\documentclass[10pt,journal,compsoc]{IEEEtran-new}
\usepackage[T1]{fontenc}

\ifCLASSINFOpdf
\else
\fi
\usepackage{amsmath}
\DeclareMathOperator{\Cov}{Cov}
\usepackage{setspace} 
\usepackage{amssymb}
\usepackage{stfloats}
\usepackage{cite}
\usepackage{ragged2e}
\usepackage[ruled,vlined,linesnumbered]{algorithm2e}
\usepackage{amsfonts}
\usepackage{mathrsfs}
\usepackage{amsmath,amsthm}
\usepackage{array,booktabs}
\usepackage{subfigure}
\usepackage{multirow}
\usepackage{cuted}
\usepackage{multicol}
\usepackage{graphicx,xcolor,bm}
\usepackage{threeparttable}
\usepackage{dcolumn}
\usepackage{setspace}
\DeclareMathOperator*{\argmax}{argmax}
\usepackage{cite,bm}
\graphicspath{{figures/}}
\allowdisplaybreaks

\begin{document}

\title{Resource Trading in Edge~Computing-enabled IoV: An Efficient Futures-based Approach}
%
%

\author{Minghui Liwang, \IEEEmembership{Member, IEEE}, Ruitao Chen, and Xianbin Wang, \IEEEmembership{Fellow, IEEE}

\thanks{Minghui Liwang, Ruitao Chen, and Xianbin Wang are with the Department of Electrical and Computer Engineering, Western University, Ontario, Canada. E-mail: $\{$mliwang, rchen328,  xianbin.wang$\}$@uwo.ca.}
}

\IEEEtitleabstractindextext{
\begin{abstract}
\justifying
Mobile edge computing (MEC) has become a promising solution to utilize distributed computing resources for supporting computation-intensive vehicular applications in dynamic driving environments. To facilitate this paradigm, the onsite resource trading serves as a critical enabler. However, dynamic communications and resource conditions could lead unpredictable trading latency, trading failure, and unfair pricing to the conventional resource trading process. To overcome these challenges, we introduce a novel futures-based resource trading approach in edge computing-enabled internet of vehicles (EC-IoV), where a forward contract is used to facilitate resource trading related negotiations between an MEC server (seller) and a vehicle (buyer) in a given future term. Through estimating the historical statistics of future resource supply and network condition, we formulate the futures-based resource trading as the optimization problem aiming to maximize the seller’s and the buyer’s expected utility, while applying risk evaluations to relieve possible losses incurred by the uncertainties in the system. To tackle this problem, we propose an efficient bilateral negotiation approach which facilitates the participants reaching a consensus. Extensive simulations demonstrate that the proposed futures-based resource trading brings considerable utilities to both participants, while significantly outperforming the baseline methods on critical factors, e.g., trading failures and fairness, negotiation latency and cost.

\end{abstract}

}

\maketitle

%
\IEEEpeerreviewmaketitle

\section{Introduction}

\IEEEPARstart{T}{he}
rapid evolution of smart vehicles facilitates the development of internet of vehicles (IoV), which offers safety, convenience, and entertainment in intelligent driving environments. Furthermore, technological advances on computing and sensing enable innovative solutions for IoV applications (e.g., 3D modeling, personalized navigation, advanced driver assistants, and online AR/VR gaming) which commonly require significant amount of computational resources~\cite{1,2}.  


However, the limited computational speed and resources of a single vehicle may be insufficient to fulfill these computation-intensive applications. To satisfy the growing computational demands while improving the experience of vehicular users, mobile edge computing (MEC) has been considered as an effective solution to IoV related applications \cite{3,4}, which brings the cloud capacity to the edge of the network, and thereby offers flexible and cost-effective computing services~\cite{5}.




The MEC service provisioning in IoV always relies on a resource trading environment, where a vehicle with heavy computational workload can offload its tasks to a nearby MEC server via vehicle-to-infrastructure (V2I) communications. To achieve this paradigm with proper incentive, conventional resource trading is utilized, enabling participants to buy or sell onsite resources for the imminent tasks \cite{6}. Specifically, participants would reach a consensus on terms such as price and the amount of resources, depending on the dynamic available resource and changing network conditions. However, this onsite trading scheme would inevitably lead to undesirable performance degradations. First, some onsite participants may face with failures to access resources. Then, the unguaranteed negotiation cost that the participants have to spend to reach the consensus on trading, may lead to unsatisfactory user experience. Additionally, onsite trading may incur significant unfairness on resource pricing owing to the inherent randomness of trading market (e.g., resource supply and demand, network conditions, etc).





The major challenges of applying onsite resource trading to IoV can be summarized as follows:

\noindent
$\bullet$ \textit{Negotiation cost}: negotiation cost mainly contains the latency, and the other cost such as energy consumption and signaling overhead, etc. Specifically, the real-time onsite negotiation will incur unnecessary trading latency, which could dramatically reduce the usable time for resource sharing. Moreover, the mobile devices that are sensitive to energy and battery bring challenges to the trading mechanism design. Under dynamic IoV environment, factors such as the limited V2I connect duration, the demands of real-time vehicular applications and the ever-changing network conditions, pose constraints to the negotiation among resource owners and requestors. To facilitate the resource trading, how to achieve the timely resource provisioning under fast-changing network environments represents one of the key challenges.

\noindent
$\bullet$ \textit{Trading failure}: a trading fails when the participants fail to reach the consensus (e.g., either of the participants gets negative utility), which would prevent the timely provisioning of expected resource, and thus lead to unsatisfactory trading experience. Therefore, recognizing and avoiding potential trading failures is indispensable for supporting ultra reliable and robust edge computing. 


\noindent
$\bullet$ \textit{Unfairness}: unfairness is mainly incurred by fluctuating prices (e.g., larger fluctuation leads to worse fairness), owing to the uncertainties such as varying V2I channel qualities and resource supply. Thus, it is imperative to alleviate or avoid the undesirable unfairness in resource trading mechanism design. 


The major motivation of this paper is to address the abovementioned challenges. We study a novel  futures-based~\cite{6} resource trading approach in edge computing-enabled IoV (EC-IoV), that considers an MEC server as the resource owner (seller) and a smart vehicle with heavy workload as the resource requestor (buyer), which could access the resources of seller via road side units (RSUs). Specifically, futures presents a forward contract where participants reach a consensus on trading a certain amount of resources in the future with predetermined price, which enables trading fairness (e.g., smooth pricing) and trading efficiency (e.g., low negotiation latency and cost), while facilitating the low-risk of trading failures. We formulate the resource trading as the optimization of both the seller's and the buyer's expected utility, and propose an efficient solution where the two participants negotiate a bilateral forward contract that predetermines the resource amount and the relevant unit price. Note that the unpredictable resource supply and network condition may lead to unsatisfactory utilities during the fulfillment of the forward contract, we apply risk evaluations to relieve possible losses for both the participants.




\begin{figure*}[h!t]
\centerline{\includegraphics[width=0.98\linewidth]{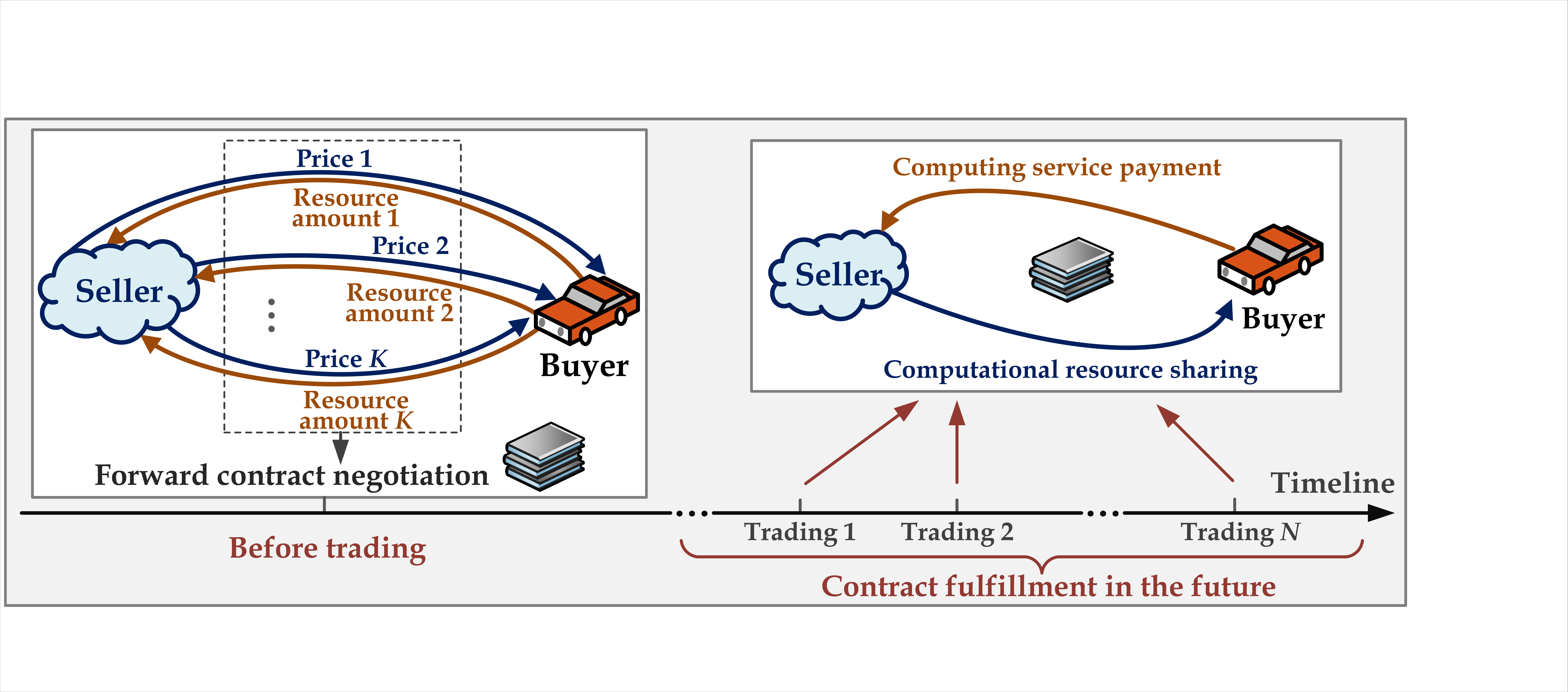}}
\caption{Framework of futures-based resource trading in EC-IoV: the seller and the buyer negotiate a forward contract on the amount of resources and the relevant price before trading, and the contract will be fulfilled in the future. The futures-based approach brings trading fairness and negotiation efficiency, while reducing the risk of trading failure.}
\label{fig1}
\end{figure*}

%


\subsection{Related work}

The existing works devoted to studying resource trading can be roughly divided into two categories: the onsite trading where participants reach an agreement relying on the current network conditions (e.g., onsite game~\cite{7,8,9,10} and auction~\cite{11,12,13,14}); and the futures-based trading, where participants sign a forward contract over buying or selling a certain number of commodities at a reasonable price in the future. For the onsite trading market, \textit{LiWang et al}.\cite{7} introduced a Stackelberg game based opportunistic computation offloading approach among moving vehicles, where the equilibriums were provided under complete and incomplete information environments. \textit{Liu et al}.\cite{8} tackled a multi-user game-based computation offloading problem in vehicular edge networks via a distributed algorithm to obtain Nash equilibrium.
 \textit{Wang et al}.\cite{9} proposed a multi-user non-cooperative offloading game, aiming to maximize the utility of each vehicle via a distributed best response algorithm.
In~\cite{10}, \textit{Jo\v{s}ilo et al}. developed a game theory model for the computation offloading problem for autonomous devices, aiming at minimizing their cost. A truthful double auction mechanism was proposed by \textit{Lu et. al.} in~\cite{11} to bridge users' task requirements and providers' resources in two-sided cloud markets. In~\cite{12}, \textit{Gao et al}. modeled the VM resource allocation problem among edge clouds and mobile users as an n-to-one weighted bipartite graph matching problem with 0-1 knapsack constraints, and designed a greedy approximation algorithm. \textit{Wang et al}.\cite{13} investigated a distributed auction model to facilitate the resource trading between the owner of the tasks and the mobile devices participating in task execution. \textit{Gao et al}.\cite{14} proposed a truthful auction for the computation resource trading market dealing with graph tasks. However, in an onsite trading market, participants may suffer from trading failures, which lead to unsatisfactory user experience. Besides, the random nature of networks usually brings inevitable unfairness, as well as the heavy latency and cost for reaching the consensus, and thus poses challenges to the resource trading mechanism design. 

Consequently, the futures-based trading~\cite{6} has been desirably applied in financial and commodity exchange market, enabling the considerable implementation in reducing the risks of fluctuant prices and failures, as well as heavy negotiation cost. Existing studies mainly investigated the electricity market~\cite{15,16,17,18}, spectrum trading~\cite{19,20,21}, and grid computing~\cite{22}. \textit{Khatib et al}.\cite{15} studied a mutually beneficial and risk tolerable forward bilateral contract in mixed pool/bilateral electricity markets. In~\cite{16}, \textit{S\'{a} nchez et al}. provided results that cast doubt on the assumption about if introducing voluntary forward markets will mitigate the market power of electricity generating companies, via encouraging them to sign a forward contract. \textit{Wang et al}.\cite{17} introduced an optimal dynamic hedging of electricity futures using copula-garch models. In~\cite{18}, \textit{Morales et al}. proposed a scenario reduction procedure that advantageously compared with the existing ones for electricity-market problems via two-stage stochastic programming, for futures market trading. As for spectrum trading, \textit{Li et al}.\cite{19} investigated a futures market to manage the financial risk and discover future spectrum price. A hybrid market approach of spectrum trading was proposed by \textit{Gao et al}. in~\cite{20}, where the optimal offline policy and the online Vickrey-Clarke-Groves auction were applied for the futures and spot markets, respectively. \textit{Sheng et al}.\cite{21} introduced a futures-based spectrum trading scheme to tackle the risk of trading failure and unfairness. Consider the grid resource management, \textit{Vanmechelen et al}.\cite{22} proposed a hybrid market in which a low-latency spot market coexists with a higher latency futures market. Nevertheless, few of the studies paid attention to the computational resources trading in MEC-assisted networks. To the best of our knowledge, this paper is  among the first to study the efficient futures-based trading mechanism under EC-IoV framework.


%
\subsection{Novelty and contributions}

In this paper, we present a futures-based resource trading approach considering an MEC server as the seller; and a smart vehicle as the buyer.
Specifically, the two participants negotiate a forward contract about the amount of resources and the relevant unit price under risk evaluation, via considering the uncertainty of resource availability and network conditions. Major contributions are summarized below:

\noindent
1). We establish a novel futures-based trading market under EC-IoV framework, allowing the buyer with heavy workload to offload its computation-intensive tasks to the seller with available resources, while achieving trading fairness and efficiency. Particularly, the two participants are motivated to reach a forward contract, which will be fulfilled in the future.

\noindent
2). We formulate the seller's utility concerning the dynamism of the number of local users, and the possible waiting cost incurred by selling superabundant resources. Moreover, we define the buyer's utility as a trade-off among the task execution time, the payment for computing service, and the transmission delay for offloading the relevant task data, that is susceptible to the uncertain V2I channel conditions (e.g., impacted by the distance between the vehicle and RSU).
The objectives of the forward contract are formulated as the maximization of both the seller's and the buyer's expected utility, while evaluating the risks of sustaining possible losses.

\noindent
3). To tackle the forward contract, we propose a bilateral negotiation approach that contains two key steps: price presetting, and iterative negotiation; through which, the participants can efficiently reach the consensus on the amount of resources and the relevant reasonable price. 

\noindent
4). Based on thorough numerical analysis and comparative evaluations, we demonstrate that the proposed futures-based trading enables commendable utilities for both the seller and the buyer, and achieves significant improvements on indicators such as trading failures, average buyer attrition rate, negotiation latency and cost, as well as fairness, against the baseline methods (onsite trading, and future-based trading without risk evaluation).

The rest of this paper is organized as follows. In Section 2, we present the framework of the proposed futures-based resource trading, as well as the models of the seller and buyer. We formulate the resource trading problem, and propose a novel bilateral negotiation mechanism for the forward contract in Section 3. Numerical results and performance evaluation are introduced in Section 4, before drawing the conclusion and future work in Section 5.

\begin{table}[htb]
{\small
\begin{center}
\caption{Major notations}
\begin{tabular*}{\linewidth}{cl}
\toprule
Notation& 
\multicolumn{1}{c}{Explanation} \\
\hline
\\
$\mathcal{A}$, $\mathcal{P}$ & 
The amount, and unit price of resources \\
$M$ & 
The total number of VMs of the seller \\
$U^{s}$, $ C^{s}$ & 
Local revenue, the cost incurred by trading\\
$n_l$, $ p_l$ & 
The number of local users, unit local revenue \\
$U^{b}$& 
The total saved execution time\\
$ C^{b}$ &
The data transmission delay\\
${\gamma}_{b, s}$, $d$ & 
 SNR of V2I link, the data size of a task\\
$\mathcal{W}$& 
The bandwidth of V2I link \\
$\tau$ &
The saved execution time per task\\
$\mathrm{E[]}$& 
Expectation of random variable\\
$\rm Ei ()$ &
Exponential integral function \\
$p^{max}_{s}$, $ p^{min}_{s}$ & 
The maximum/minimum price of the seller\\
$ \Delta p$ &
The adjustment granularity of the price \\
$U^{b}$& 
The total saved execution time\\
${\mathcal{U}}^{s} (n_l, \mathcal{A}, \mathcal{P})$ & 
Utility of the seller\\
$\overline{{\mathcal{U}}^{s}}(n_l, \mathcal{A}, \mathcal{P})$ & 
Expected utility of the seller \\
${\mathcal{R}}^{s}(n_l, \mathcal{A}, \mathcal{P})$ & 
Risk of the seller \\
${\mathcal{U}}^{b}({\gamma}_{b, s}, \mathcal{A}, \mathcal{P})$ & 
Utility of the buyer\\
$\overline{{\mathcal{U}}^{b}}({\gamma}_{b, s}, \mathcal{A}, \mathcal{P})$  & 
Expected utility of the buyer\\
${\mathcal{R}}^{b}({\gamma}_{b, s}, \mathcal{A}, \mathcal{P})$  & 
Risk of the buyer\\
\bottomrule
\end{tabular*}
\label{tab1}
\end{center}
}
\end{table}

\section{System Model}
\subsection{Framework of futures-based resource trading in EC-IoV}

The framework of the proposed futures-based resource trading in EC-IoV is given in {\color{black}Fig.}~1, which contains two participants: the seller (MEC server) and the buyer (vehicle). The seller owns a collection of machines~\cite{23,24} where each machine has a set of virtual machines (VMs) that can run a certain number of tasks in parallel. Specifically, VMs and tasks are matched to each other following one-to-one mode. A local user is seen as a regular customer of the seller, e.g., the annual membership of the computing service. Notably, each local user has a task and being collected in a buffer waiting to be processed. The buyer has a set of tasks\footnote{In this paper, assume that the buyer has a a large amount of tasks need to processed, and we do not consider the exact number of tasks in this set, which will be investigated in our future work.} requiring to be processed, which, however, suffers from the insufficient on-board computational resources and limited capability.
In consequence, the buyer tends to trade with the seller by paying for a certain amount of VMs for task execution. 

Notably, the privacy of information between participants poses challenges to the resource trading mechanism design (the seller and the buyer do not know the information of each other, e.g., the number of seller's local users is unknown to the buyer). Moreover, the contract period is out of the scope of this paper; namely, the seller and the buyer can negotiate another forward contract when the previous contract is about to expire. 

This paper mainly studies how the two participants determine the amount of VMs $\mathcal{A}$ ($\mathcal{A} \ge 1$, $\mathcal{A}=0$ when a trading fails) and the relevant unit price $\mathcal{P}$ via a mutually beneficial forward contract for future resource trading.

\subsection{Modeling of the seller: utility and risk}

\begin{figure}[t!]
\centering
\subfigure[]{\includegraphics[width=.48\linewidth]{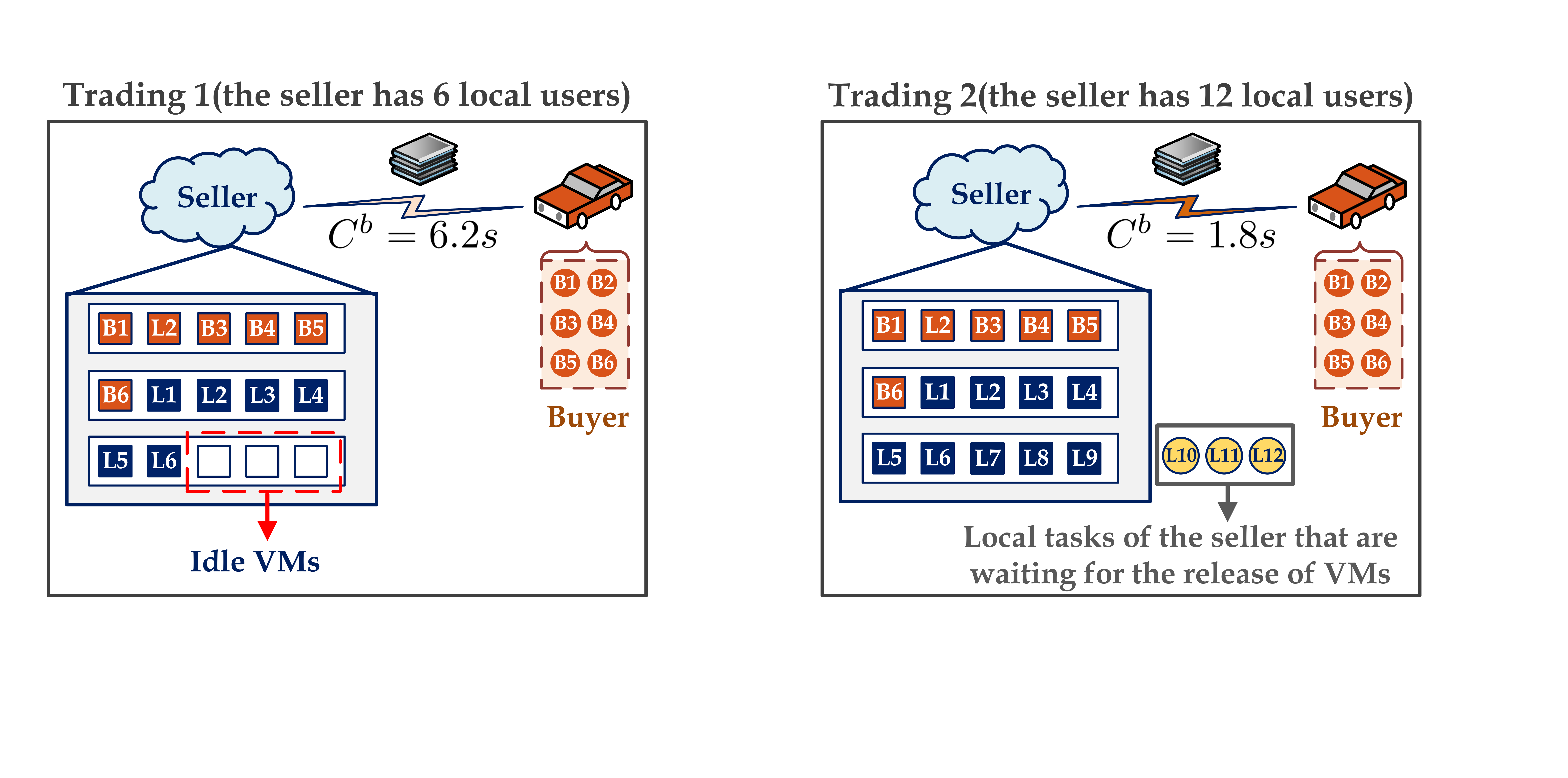}}
\subfigure[]{\includegraphics[width=.48\linewidth]{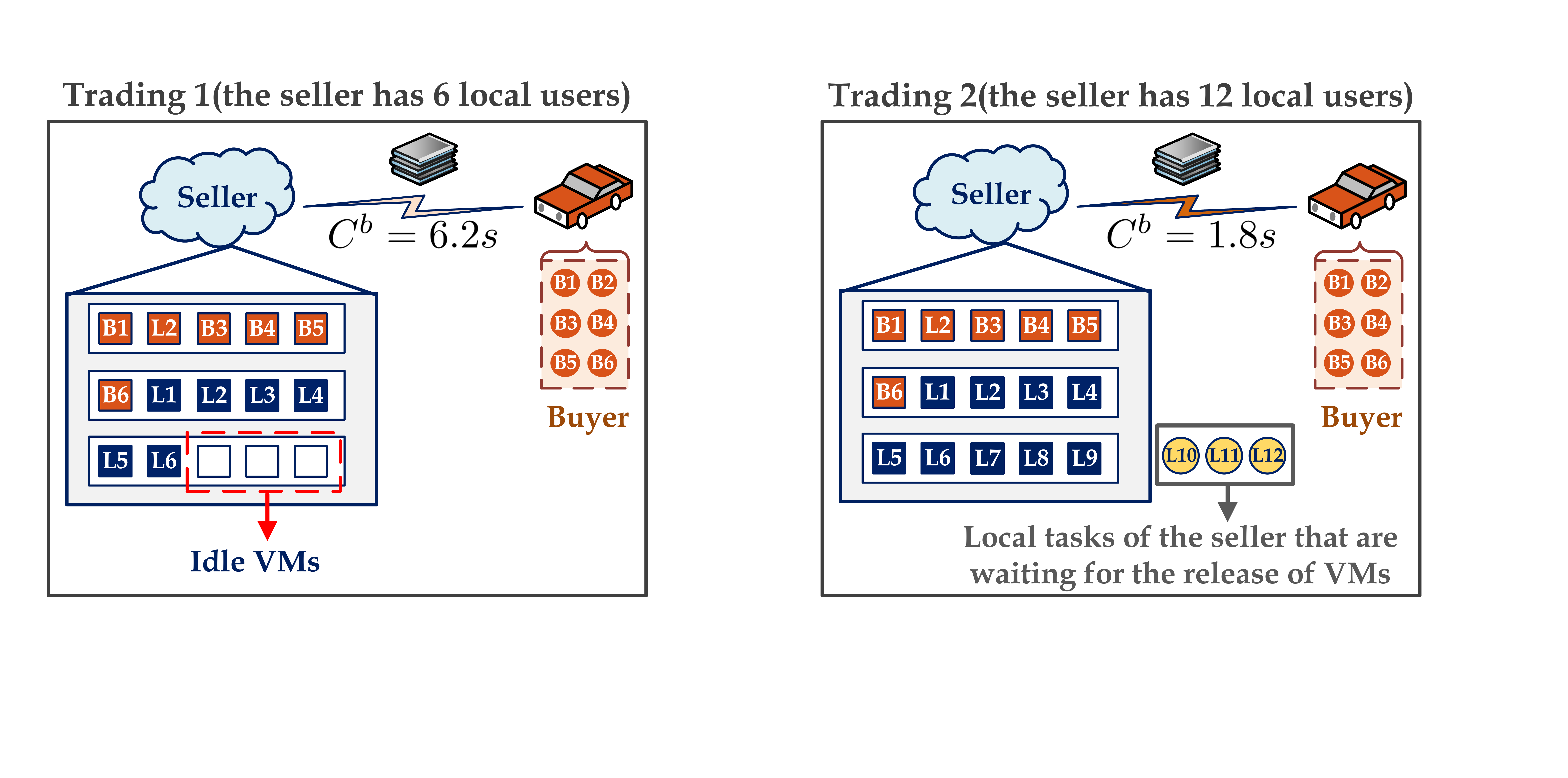}}
\caption{Examples of the main uncertainties ($n_l$ and $\gamma_{b,s}$), where the amount of trading resources in the forward contract is 6, and  $\gamma_{b,s}$ in trading 1 is worse than that in trading 2: (a). $n_l=6$, $C^s=0$; (b). $n_l=12$, $C^s=3\times r_l$.}
\end{figure}
In this section, we introduce the utility and risk of the seller. Let the total number of the available VMs provided by the seller be $M$ ($M\ge \mathcal{A}$).

\textbf{Seller's utility:} The utility of the seller mainly contains the revenue from local users $U^{s}$, the income obtained from trading with the buyer $\mathcal{A}\times\mathcal{P}$, and the cost $C^{s}$ incurred by the trading, which is defined as (1).
\begin{align}
\label{eq1}
{\mathcal{U}}^{s}(n_l, \mathcal{A}, \mathcal{P})=U^{s}+\mathcal{A}\times \mathcal{P}-C^{s},
\end{align}
where $U^{s}=n_l\times p_l$ and $n_l\in \{0,1,2,\dots, M\}$ indicates the number of local users, following a discrete uniform distribution. $p_l$ presents the unit revenue for serving a local user. The cost $ C^{s}$ is defined as the refund to the local users that have to wait for the release of occupied VMs:
\begin{align}
\label{eq2}
C^{s}=\begin{cases}
0,& {0\le n}_l\le M-\mathcal{A} \\
c_l(n_l-(M-\mathcal{A})), & M-\mathcal{A}< n_l\le M
\end{cases},
\end{align}

where $c_l$ ($c_l\le p_l$) denotes the unit waiting cost owing to the factor that the remaining VMs after resource trading cannot meet the current local task requirements. Namely, a part of the local tasks will be processed after the release of VMs when $ n_l>M-\mathcal{A}$. Otherwise, $C^{s}=0$. Examples are shown in Fig. 2, and {\color{black}Fig.}~3(a) describes the variation tendency of ${\mathcal{U}}^{s}$ with increasing value of $n_l$.

One of the key concerns for participants in this paper is to make a contract, based on which, the relevant resource trading will be realized in the future. However, owing to the future's uncertainty (e.g., the unpredictable number of local users), the buyer and the seller are responsible for their own profits and risks of loss when fulfilling the contract. In consequence, the seller provides an acceptable tolerance of the risk to reach the agreement of the futures contract with the buyer, during negotiation. Particularly, the risk of the seller mainly derives from the prediction uncertainty of $n_l$, where a larger $n_l$ leads to heavier cost. Thus, we define the seller's ideal trading condition as (3):
\begin{align}
\label{eq3}
{\mathcal{U}}^{s}(n_l, \mathcal{A}, \mathcal{P})>{\lambda} ^{s}_1\times \overline{{\mathcal{U}}^{s}}(n_l, \mathcal{A}, \mathcal{P}),
\end{align}

where $\overline{{\mathcal{U}}^{s}}(n_l, \mathcal{A}, \mathcal{P})={\rm E}\left[{\mathcal{U}}^{s}(n_l, \mathcal{A}, \mathcal{P})\right]$ indicates the expected utility of the seller, ${\lambda} ^{s}_1$ denotes a threshold coefficient. The inequation (3) describes the condition where the seller always tends to achieve a better utility than its expectation. Specifically, $\overline{{\mathcal{U}}^{s}}(n_l, \mathcal{A}, \mathcal{P})$ is computed as (4),
\begin{align}
\label{eq4}
\overline{{\mathcal{U}}^{s}}(n_l, \mathcal{A}, \mathcal{P})=p_l{\rm E}[n_l]+\mathcal{A}\times \mathcal{P}-{\rm E}[C^{s}],
\end{align}

where ${\rm E}[n_l]={M}/{2}$. We calculate the expectation of the cost $C^{s}$ as ${\rm E}[C^{s}]={{c_l\mathcal{A}}^2+c_l\mathcal{A}}/{2 (M+1)}$, the detailed derivation of which is given in \textbf{Appendix}. Thus, $\overline{{\mathcal{U}}^{s}}(n_l, \mathcal{A}, \mathcal{P})$ is further given as (5), which indicates a monotone increasing function of $\mathcal{P}$ under any given $\mathcal{A}$ ($\mathcal{A}\neq 0$); and a quadratic equation of $\mathcal{A}$ given any unit price $\mathcal{P}$. The diagram of which is shown in {\color{black}Fig.}~3(c).
\begin{align}
\label{eq5}
\overline{{\mathcal{U}}^{s}}(n_l, \mathcal{A}, \mathcal{P})=\frac{Mp_l}{2}+\mathcal{A}\times \mathcal{P}-\frac{{c_l\mathcal{A}}^2+c_l\mathcal{A}}{2 (M+1)}
\end{align}

\textbf{Seller's risk:} Correspondingly, we define the risk of the seller ${\mathcal{R}}^{s}(n_l, \mathcal{A}, \mathcal{P})$ as the probability that the ratio of ${\mathcal{U}}^{s}(n_l, \mathcal{A}, \mathcal{P})$ and the expectation $\overline{{\mathcal{U}}^{s}}(n_l, \mathcal{A}, \mathcal{P})$ is less than the threshold ${\lambda} ^{s}_1$, shown in (6).

\begin{align}
\label{eq6}
{\mathcal{R}}^{s}(n_l, \mathcal{A}, \mathcal{P})={\rm Pr}\left\{\frac{{\mathcal{U}}^{s}(n_l, \mathcal{A}, \mathcal{P})}{\overline{{\mathcal{U}}^{s}}(n_l, \mathcal{A}, \mathcal{P})}\le {\lambda}^{s}_1\right\}
\end{align}

Combine with (5), (6) is thus rewrote as (7).
\begin{align}
\label{eq7}
{\mathcal{R}}^{s}(n_l, \mathcal{A}, \mathcal{P})={\rm Pr}\{U^{s}-C^{s}\le {\lambda} ^{s}_1\times \overline{{\mathcal{U}}^{s}}(n_l, \mathcal{A}, \mathcal{P})-\mathcal{A}\times \mathcal{P}\}
\end{align}
For notational simplicity, let discrete random variable $S=U^s-C^s$ represent the left side of ``$\le $'' in (7), which stands for a piece-wise function of $n_l$ given in (8),
\begin{align}
\label{eq8}
\hspace{-1.1em}S=\begin{cases}
p_ln_l, 0\le n_l\le M-\mathcal{A} \\
(p_l-c_l)n_l+c_l(M-\mathcal{A}), M-\mathcal{A}+1\le n_l\le M,
\end{cases}
\end{align}
\noindent
where $r$ denotes the right side of ``$\le $'' in (7), given by (9).
\begin{align}
\label{eq9}
r&={\lambda}^s_1\times \overline{\mathcal{U}^s}(n_l, \mathcal{A}, \mathcal{P})-\mathcal{A}\times \mathcal{P}\notag \\
&=- \frac{{\lambda}^{s}_{1}{{c}_{l}\mathcal{A}}^{2}+{{\lambda}^{s}_{1}c}_{l}\mathcal{A}}{2(M+1)}+({\lambda}^{s}_{1}-1)\mathcal{P}\times \mathcal{A}+\frac{{\lambda}^{s}_{1}M{p}_{l}}{2}
\end{align}
Thus, we have the probability mass function (PMF) of the discrete random variable $S$ as (10).
\begin{align}
\label{eq10}
& {\rm Pr}(S=k)=\frac{1}{M+1}, k\in \{0,p_l,2p_l, \dots , p_l(M-\mathcal{A}), \notag\\
&  p_l(M \!-\! \mathcal{A})\!+\! (p_l \!-\! c_l), p_l(M \!-\! \mathcal{A})\!+\! 2(p_l \!-\! c_l), \dots, p_lM \!-\! c_l\mathcal{A} \}\tag{10}
\end{align}
Accordingly, the risk ${\mathcal{R}}^{s}({n}_{l}, \mathcal{A}, \mathcal{P})$ is rewritten as (11) via calculating the cumulative distribution function (CDF) of $S$, where $\lfloor \cdot  \rfloor $ stands for the round down operation (find derivations of (10) and (11) in \textbf{Appendix}). Apparently, a higher risk may lead to a lower utility of the seller. 

\begin{strip}
\hrulefill
\begin{align}
\label{eq11}
& {\mathcal{R}}^{s}({n}_{l}, \mathcal{A}, \mathcal{P})={\rm Pr}\{S\le r\}=
\begin{cases}
0, & r<0\\ 
\dfrac{\left\lfloor \frac{r}{p_l}\right\rfloor +1}{M+1}, & 0\le r< {p}_{l}(M-\mathcal{A})+({p_l}-{c_l}) \\
\dfrac{M-\mathcal{A}+1}{M+1}+\dfrac{\left\lfloor\frac{r-{p}_{l}(M-\mathcal{A})}{({p_l}-{c_l})}\right\rfloor}{M+1}, & {p}_{l}(M\!-\!\mathcal{A})\!+\! ({p_l}\!-\!{c_l})\!\le\! r\!\le\! {p_l}M-{c_l}\mathcal{A} \\
1, & r>{p}_{l}M-{c_l}\mathcal{A}\tag{11}
\end{cases}
\end{align}
\hrulefill
\end{strip}

\subsection{Modeling of the buyer: utility and risk}

\textbf{Buyer's utility:} The utility of the buyer ${\mathcal{U}}^{b}({\gamma}_{b, s}, \mathcal{A}, \mathcal{P})$  consists of the saved execution time $U^{b}$ from enjoying the computing service, the payment $\mathcal{A}\times \mathcal{P}$ for the trading resources, and the data transmission delay $C^{b}$ for offloading tasks to the seller. Namely, the buyer has to transfer the corresponding task data to the seller through V2I communications\footnote{We particularly concern the execution time and uplink transmission delay of tasks, while ignoring the downlink transmission delay for result feedback owing to the noncomparable output data size~\cite{9}.}. Correspondingly, ${\mathcal{U}}^{b}({\gamma}_{b, s},\mathcal{A},\mathcal{P})$ is defined as
\begin{align}
\label{eq12}
{\mathcal{U}}^{b}({\gamma}_{b, s}, \mathcal{A}, \mathcal{P})=U^{b}-\omega\mathcal{A}\times \mathcal{P}-C^{b},\tag{12}
\end{align}
where $\omega $ denotes a non-negative weight coefficient to balance different measure units (e.g., time and payment). We define $U^b$ as $U^{b}=\mathcal{A}\times \tau $, where $\tau $ presents the saved execution time per task. $C^{b}$ is calculated by (13), 
\begin{align}
\label{eq13}
C^{b}=\frac{\mathcal{A}\times d}{\mathcal{W}\log_2(1+{\gamma}_{b, s})},\tag{13}
\end{align}
where ${\gamma}_{b,s}$ denotes the signal-to-noise ratio (SNR) of the V2I communication link between the buyer and the seller, which is a random variable owing to the factors such as the uncertainties of wireless communication environment, and the distance between the buyer and the current accessed RSU\footnote{In this paper, assume the vehicle can get access to the seller via a RSU during each trading, and thus the exact location and mobility of the vehicle are not necessarily considered.}. Suppose that ${\gamma}_{b,s}$ follows an uniform distribution~\cite{21} in interval $[{\varepsilon}_1,{\varepsilon}_2]$, denoted by ${\gamma}_{b, s} \sim {\rm U} ({\varepsilon}_1,{\varepsilon}_2)$, where ${\varepsilon}_1$ and ${\varepsilon}_2$ are positive parameters. $d$ (bit) indicates the data size of a task\footnote{In this paper, assume that all the tasks have the same data size. However, we can also consider the cases with different data sizes of tasks, for which our proposed approach can still be well applied.}, and $\mathcal{W}$ represents the related channel bandwidth. Examples of the uncertainty of $\gamma_{b,s}$ are given in Fig. 2; and {\color{black}Fig.}~3(b) presents the diagram of the variation tendency of ${\mathcal{U}}^{b}$ when considering $\tau -\omega\mathcal{P}>0$.   

\textbf{Buyer's risk:} To alleviate the heavy on-board workloads, the buyer is always willing to trade with the seller when ${\mathcal{U}}^{b}>0$. Nevertheless, a poor channel quality (e.g., a small value of ${\gamma}_{b, s}$) may lead to a negative value of ${\mathcal{U}}^{b}$. Thus, let the minimum of ${\mathcal{U}}^{b}$ be ${\mathcal{U}}^{b}_{min}$, which is defined as a value approaches to zero (e.g., ${\mathcal{U}^{b}_{min}=10^{-8}}$), to protect the task execution requirements of the buyer. Correspondingly, the risk ${\mathcal{R}}^{b}({\gamma}_{b, s}, \mathcal{A}, \mathcal{P})$ of the buyer is mainly incurred by the prediction uncertainty of the randomness of SNR ${\gamma}_{b, s}$, which is defined as the probability that ${\mathcal{U}}^{b}$ may be too close to ${\mathcal{U}}^{b}_{min}$.
\begin{align}
\label{eq14}
{\mathcal{R}}^{b}({\gamma}_{b, s}, \mathcal{A}, \mathcal{P})={\rm Pr}\left\{\frac{{\mathcal{U}}^{b}({\gamma}_{b, s}, \mathcal{A}, \mathcal{P})}{{\mathcal{U}}^{b}_{min}}\le {\lambda} ^{b}_1\right\}\tag{14}
\end{align}
Similarly, the buyer also has a risk tolerance to accept the futures contract, where ${\lambda}^{b}_1$ in (14) represents a threshold coefficient. Combine with (12) and (13), (14) can be further written as (15),
\begin{align}
\label{eq15}
{\mathcal{R}}^{b}({\gamma}_{b, s}, \mathcal{A}, \mathcal{P})={\rm Pr}\left\{{\gamma}_{b, s}\le 2^{r'}-1\right\},\tag{15}
\end{align}
where $r'=\dfrac{\mathcal{A}\times d}{\mathcal{W}(\mathcal{A}\times \tau-{\mathcal{U}}^{b}_{min}{\lambda} ^{b}_1-\omega \mathcal{A}\times \mathcal{P})}$. According to ${\gamma}_{b, s} \sim {\rm U} ({\varepsilon}_1,{\varepsilon}_2)$, ${\mathcal{R}}^{b}({\gamma}_{b, s}, \mathcal{A}, \mathcal{P})$ is calculated as (16).
\begin{align}
\label{eq16}
& {\mathcal{R}}^{b}({\gamma}_{b, s}, \mathcal{A}, \mathcal{P})=\notag \\
& \begin{cases}
0, & r'<\log_{2}({\varepsilon}_1+1) \\
\dfrac{2^{r'}-{\varepsilon}_1-1}{{\varepsilon}_2-{\varepsilon}_1}, & \log_{2}({\varepsilon}_1+1)\le r'\le \log_{2}({\varepsilon}_2+1) \\
1, & r'>\log_{2}({\varepsilon}_2+1)\tag{16}
\end{cases}
\end{align}
Moreover, the buyer's expected utility $\overline{{\mathcal{U}}^{b}}({\gamma}_{b, s}, \mathcal{A}, \mathcal{P})={\rm E}\left[{\mathcal{U}}^{b}({\gamma}_{b, s}, \mathcal{A}, \mathcal{P})\right]$ is given by (17), which stands for a monotone function with coordinate $(0,0)$ when $\mathcal{P}$ is fixed; and a monotone decreasing function of $\mathcal{P}$ under any given $\mathcal{A}$. {\color{black}Fig.}~3(d) shows the diagram of $\overline{{\mathcal{U}}^{b}}({\gamma}_{b, s}, \mathcal{A}, \mathcal{P})$.
\begin{align}
\label{eq17}
& \overline{{\mathcal{U}}^{b}}({\gamma}_{b, s}, \mathcal{A}, \mathcal{P})\notag \\
& =\mathcal{A}\times \tau-\omega \mathcal{A}\times \mathcal{P}-\frac{\mathcal{A}\times d}{\mathcal{W}}\times {\rm E}\left[\frac{1}{\log_2(1+{\gamma}_{b, s})}\right]\tag{17}
\end{align}
Specifically, ${\rm E}\left[\dfrac{1}{\log_2(1+{\gamma}_{b, s})}\right]$ is given in (18) (find detailed derivation in \textbf{Appendix}), where $r''=\dfrac{\ln 2}{{\varepsilon}_2-{\varepsilon}_1}$ denotes a constant for notational simplicity; ${\rm Ei}(y)=\displaystyle\int^y_{-\infty} {\dfrac{e^x}{x}}{\rm d}x$ represents the exponential integral function.
\begin{align}
\label{eq18}
& {\rm E}\left[\frac{1}{\log_2(1+{\gamma}_{b, s})}\right]\notag \\
& =r''{\rm Ei}({\ln 2}\times \log_{2}({\varepsilon}_{2}+1))-r''{\rm Ei}({\ln 2}\times \log_{2}({\varepsilon}_{1}+1)) \tag{18}
\end{align}
\begin{figure}[h!]
\centering
\subfigure[]{\includegraphics[width=.455\linewidth]{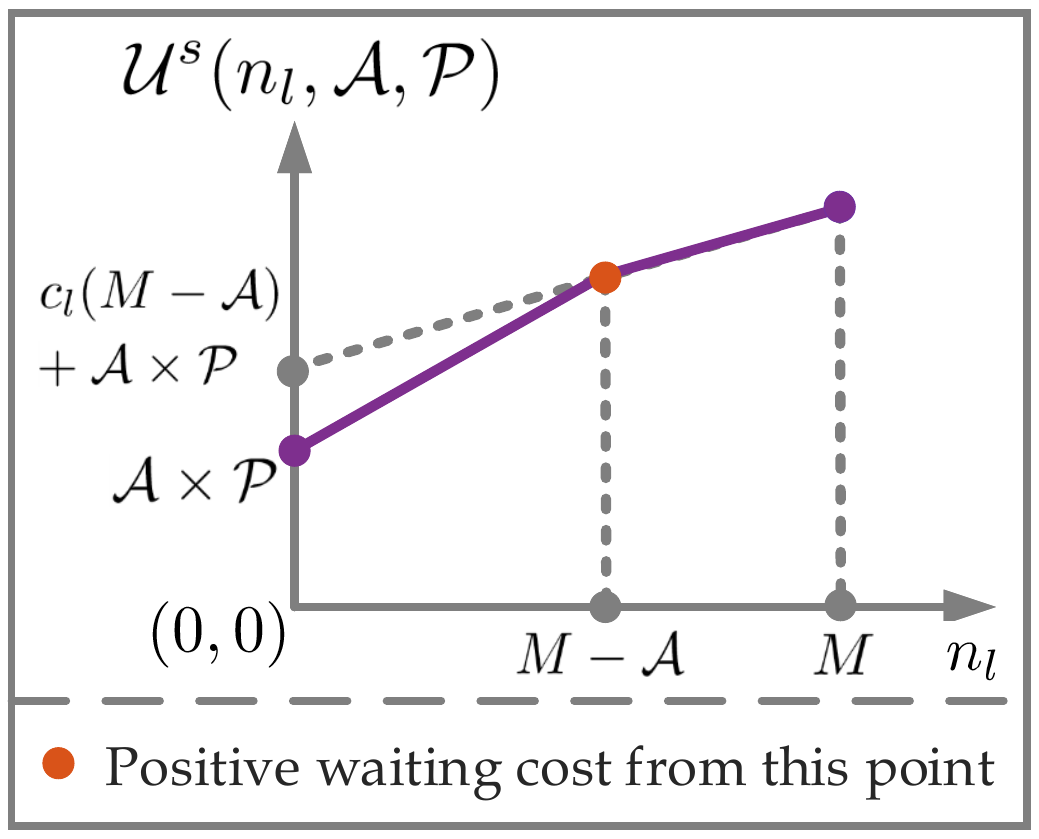}}
\subfigure[]{\includegraphics[width=.48\linewidth]{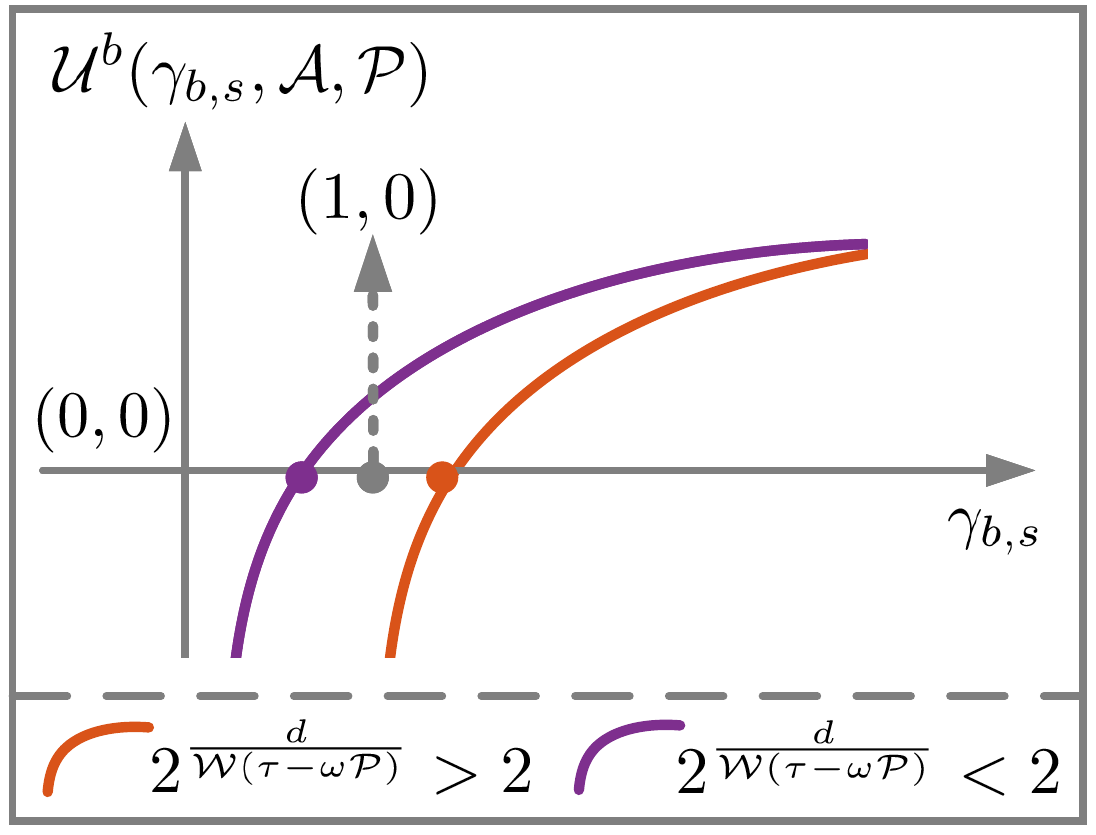}}\\
\subfigure[]{\includegraphics[width=.465\linewidth]{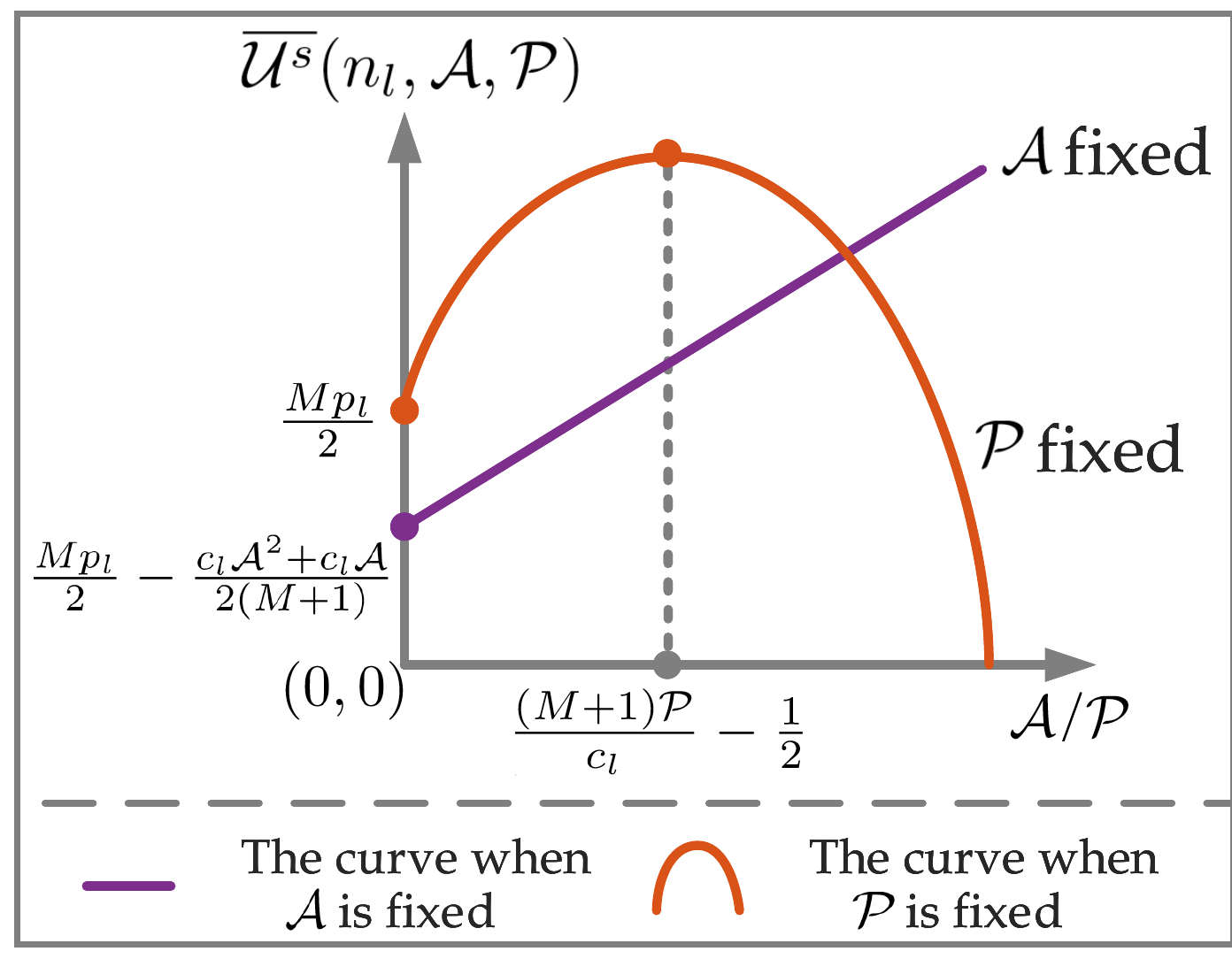}}
\subfigure[]{\includegraphics[width=.48\linewidth]{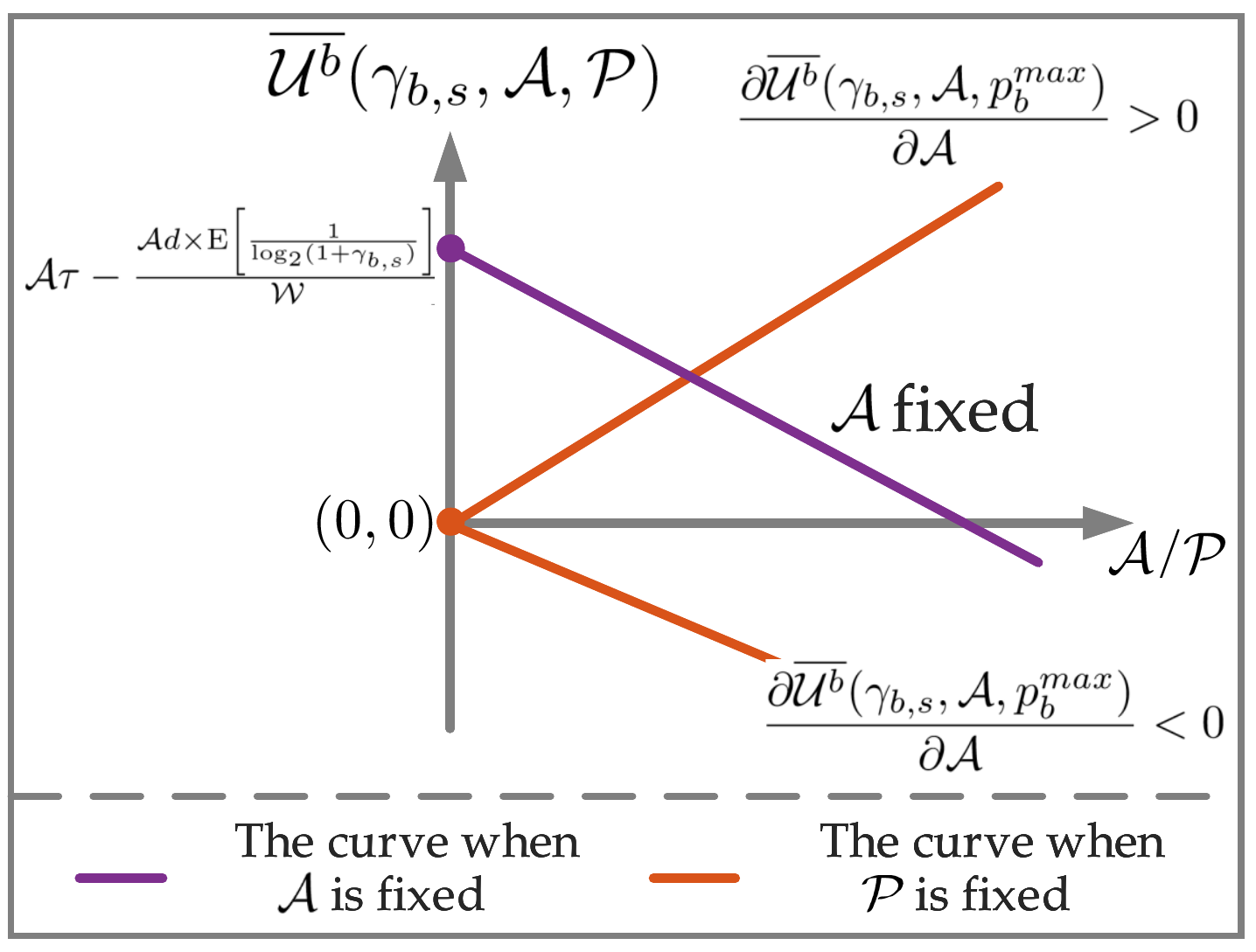}}
\caption{Diagrams of the utilities and expected utilities of participants: (a). ${\mathcal{U}}^{s}(n_l, \mathcal{A}, \mathcal{P})$; (b). ${\mathcal{U}}^{b}({\gamma}_{b, s}, \mathcal{A}, \mathcal{P})$ when $\tau-\omega \mathcal{P}>0$; (c). $\overline{{\mathcal{U}}^{s}}(n_l, \mathcal{A}, \mathcal{P})$; (d). $\overline{{\mathcal{U}}^{b}}({\gamma}_{b, s}, \mathcal{A}, \mathcal{P})$.}
\end{figure}
\subsection{Pricing criteria}
We define the pricing criteria in resource trading as (19), where $p^{max}_{s}$ and $p^{min}_{s}$ indicate the prescriptive maximum and minimum unit price of the seller, respectively. Specifically, suppose $p^{min}_{s}\ge p_l$ to distinguish the buyer with the local users\footnote{In a real-life resource trading market, the buyer is always required to pay more than the local user who owns the membership.} of the buyer. $\Delta p$ denotes the adjustment granularity of the price and $\kappa $ stands for a positive integer. Namely, the seller has to comply with the pricing rule rather than misreporting (e.g., bidding up the unit price to get more benefits from resource trading). Apparently, we have $p^{min}_{s}\le \mathcal{P}\le p^{max}_{s}$. Major notations in this paper are summarized in {\color{black}Table}~1. 
\begin{align}
\label{eq19}
p^{max}_{s}=p^{min}_{s}+\kappa \Delta p\tag{19}
\end{align}

\section{Proposed Efficient Futures-based Resource Trading Approach}

\subsection{Problem formulation}
\noindent
In the proposed futures-based resource trading environment, both the seller and the buyer aim at maximizing its expected utility rather than the utility itself, owing to the unpredictability of the number of local users (from the seller's perspective), and the SNR of V2I communication link (from the buyer's perspective). Correspondingly, we define the objective of seller as the maximization of $\overline{{\mathcal{U}}^{s}}$, while meeting the constraint of the acceptable tolerance for risk, which is formulated as optimization problem ${\mathcal{F}}^{s}$ given in (20) and (C1),
\begin{align}
\label{eq20}
{\mathcal{F}}^{s}: \max_ {\mathcal{A}, \mathcal{P}}~\overline{{\mathcal{U}}^{s}}(n_l, \mathcal{A}, \mathcal{P})\tag{20}
\end{align}
\centerline{s.t.\hfill ${\mathcal{R}}^{ s}(n_l, \mathcal{A}, \mathcal{P})\le {\lambda} ^{s}_2$,\hfill (C1)}\\
where ${\lambda}^{s}_2$ denotes the seller's acceptable threshold of the risk. Similarly, aiming to maximize the expected utility $\overline{{\mathcal{U}}^{b}}$ while satisfying the acceptable tolerance of the risk ${\lambda}^{b}_2$, the objective of the buyer is formulated as optimization problem ${\mathcal{F}}^{b}$ given in (21) and (C2):
\begin{align}
\label{eq21}
{\mathcal{F}}^{b}: {\max_{\mathcal{A}, \mathcal{P}}~\overline{{\mathcal{U}}^{b}}({\gamma}_{b, s}, \mathcal{A}, \mathcal{P})}\tag{21}
\end{align}
\centerline{s.t.\hfill ${\mathcal{R}}^{b}({\gamma}_{b, s}, \mathcal{A}, \mathcal{P})\le {\lambda} ^{ b}_2$.\hfill (C2)}

\vspace{0.5cm}
Owing to the information privacy between the two participants, traditional solutions such as weighted sum optimization and genetic algorithms~\cite{25,26} are difficult to be utilized to solve the proposed problem. Thus, we introduce an efficient bilateral negotiation mechanism to facilitate the seller and the buyer reaching the consensus on the forward contract. 


\subsection{Proposed bilateral negotiation mechanism for the forward~contract}

This section presents a bilateral negotiation mechanism containing two steps, where the two participants negotiate both the unit price and the amount of trading resource for the futures contract following an iterative manner. The relevant pseudocode is detailed in Algorithm 1.

\textbf{Step 1. Practical pricing presetting}: aiming to facilitate an efficient resource trading, the buyer will first report its tolerable unit price $p^{max}_b$ to avoid the negative utilities before the negotiation (line 1). In this market, $p^{max}_b$ is defined as a value which enables $\dfrac{\partial {\mathcal{U}}^{b}({\varepsilon}_2,\mathcal{A}, p^{max}_b)}{\partial \mathcal{A}}>0$ under ${\gamma}_{b, s}={\varepsilon}_{2}$. Namely, if~${\mathcal{U}}^{b} $ stays non-positive even when SNR ${\gamma}_{b, s}$ reaches its maximum ${\varepsilon}_{2}$, the trading fails. According to the pricing criteria (19), $p^{max}_b$ can be calculated as (22) which also conforms to the buyer's tolerable risk requiring $r'<\log_{2}({\varepsilon}_{2}+1)$ (constraint (C2)).
\begin{align}
\label{eq22}
p^{max}_b=\left\lfloor \frac{\frac{\tau} {\omega} -\frac{d}{\omega \mathcal{W}\log_2(1+{\varepsilon}_2)}-p^{min}_{s}}{\Delta p}\right\rfloor \times \Delta p+p^{min}_{s}\tag{22}
\end{align}

Notably, if $p^{max}_b<p^{min}_{s}$, the trading fails; otherwise, we adjust the practical pricing range as $p^{min}_{s}\le \mathcal{P}\le {\min}(p^{max}_b, p^{max}_{s})$, where ${\min}(p^{max}_b, p^{max}_{s})$ refers to the smaller value between $p^{max}_b$ and $ p^{max}_{s}$. For notational simplicity, let $p^{max}\triangleq {\min}(p^{max}_b, p^{max}_{s})$ denote the practicable maximum unit price in the proposed market.

\vspace{0.3cm}
\textbf{Step 2. Forward contract negotiation:} in each iteration during the negotiation, the seller first proposes a price and decides the relevant acceptable range of the amount of trading resource $\bm{A^{s}}$, while meeting the tolerable risk constraint (C1) (Algorithm1, line 6). Given the quoted price of the seller, the buyer then identifies an affordable range of the amount of trading resource $\bm{A^{b}}$ by analyzing its tolerable risk according to (C2) (Algorithm1, line 7). If these two ranges overlap, the buyer accepts the price and sets the amount of trading resource as the one that maximizes this expected utility $\overline{{\mathcal{U}}^{b}}({\gamma}_{b, s}, \mathcal{A}, \mathcal{P})$ (Algorithm1, lines 9), and the pair of price and amount can be seen as a candidate contract term (Algorithm1, line 9). The seller updates this price to $p^{max}-\Delta p$, and starts the next iteration (Algorithm1, lines 10--11). When all the pairs of acceptable unit price and the relevant amount of trading resource are found, the seller decides $\mathcal{P}^*$ and $\mathcal{A}^*$ for the final forward contract by choosing the pair that maximizes its expected utility $\overline{{\mathcal{U}}^{s}}(n_l, \mathcal{A}, \mathcal{P})$ (Algorithm1, lines 13--14). Otherwise, if there is no available candidate term, the contract fails. Notably, it is meaningless to consider the failures of a futures contract in this paper although the trading may fail owing to factors such as a small expectation of $\gamma_{b,s}$, and large values of $\lambda^s_2$ and $\lambda^b_2$, where the onsite trading may also risk a failure.

\begin{algorithm}
\caption{The proposed bilateral negotiation mechanism for the futures-based resource trading}
\SetKwInOut{Input}{Input}\SetKwInOut{Output}{Output}
\Input{$M$, ${\varepsilon}_{1}$, ${\varepsilon}_{2}$, $\tau$, $\omega$, $d$, $\mathcal{W}$, $\Delta p$, $p_l$, $c_l$}

\Output{resource amount $\mathcal{A}^*$, and unit price $\mathcal{P}^*$ for the forward contract}


//Price presetting: the buyer calculates $p^{max}_b$, the seller resets the practicable pricing range as $[p^{min}_{s}, p^{max}]$, \newline
\mbox{}~~Tolerable risks setting: the seller and the buyer decide the tolerable risk thresholds ${\lambda} ^{s}_2$ and ${\lambda} ^{b}_2$.

//Bilateral negotiation

the seller sets $\mathcal{P}\leftarrow p^{max}$, \\
the candidate trading term set $\mathbb{C}\leftarrow \emptyset$,
$i\leftarrow 1$,

\For{$\mathcal{P}\ge p^{min}_{s}$}{

the seller decides the relevant range of the amount of trading resource ${\bm{A}}^{\bm{s}}$ based on $\mathcal{P}$, while meeting (C1),

the buyer decides the acceptable range of the amount of trading resource ${\bm{A}}^{\bm{b}}$ based on $\mathcal{P}$, while meeting (C2),

\If{${\bm{A}}^{\bm{s}}\cap {\bm{A}}^{\bm{b}}\neq \varnothing$}{

$\mathcal{A}^i\leftarrow \mathop{\argmax}\limits_{\mathcal{A}}  {\overline{\mathcal{U}^b}(\gamma_{b,s},\mathcal{A},\mathcal{P})}$,  $\mathcal{A} \in \bm{A^s}\cap \bm{A^b}$,  
$\mathbb{C}\leftarrow\mathbb{C} \cup \left\{\mathcal{A}^i, \mathcal{P}\right\} $,  
}


$\mathcal{P}\leftarrow p^{max}-\Delta p$,

$i\leftarrow i+1$,

}

\If{$\mathbb{C} \neq \emptyset$}{
$\mathcal{A}^\prime, \mathcal{P}^\prime \leftarrow \mathop{\argmax}\limits_{\mathcal{A}, \mathcal{P}}  {\overline{\mathcal{U}^s}(n_l,\mathcal{A},\mathcal{P})}$,  $\left\{\mathcal{A}, \mathcal{P}\right\} \in \mathbb{C}$,   

${\mathcal{P}^*\leftarrow \mathcal{P}}^{\prime}$, $\mathcal{A}^*\leftarrow {\mathcal{A}}^{\prime}$, \% the futures contract signed successfully with term $\left\{\mathcal{A}^*,\mathcal{P}^*\right\}$
}
\Else{
the futures-based resource trading fails,}

\textbf{end negotiation}

\end{algorithm}

\section{Simulation and Performance Evaluation}
This section presents numerical results that illustrate the validity of the proposed futures-based resource trading approach (abbreviate to ``Proposed FuturesT'' for simplicity). 

\subsection{Simulation parameter settings, baseline methods, and critical indicators}

The major parameters in this simulation are set as follows: $p_l\in \mathrm{[0.5,0.6]}$, $ c_l\in [0.4,0.5]$, $d\in [6,7]$ Mb, $\mathcal{W}\in [5,6]$ MHz, ${\varepsilon}_{1}=10$ dB, ${\varepsilon}_{1}=23$ dB, ${\lambda}^{s}_{1}\in \mathrm{[0.95,1)}$, ${\lambda}^{s}_{2}={\lambda}^{b}_{2}\in [0.25,0.4]$. The performance of the proposed approach is compared with two baseline methods listed below. Notably, a futures-based participant fulfills the trading according to the forward contract, which facilitates a low-risk situation of trading failure. However, the onsite trading may face with failures when either participant's utility is negative. Correspondingly, let ${\mathcal{U}}^{s}={\mathcal{U}}^{b}=\mathcal{A}=\mathcal{P}\triangleq 0$ when a trading fails.


\noindent
\textbf{$\bullet$ Onsite trading (abbreviate to ``OnsiteT'')}: In each trading, the participants negotiate for a consensus on the unit price and resource amount aiming to maximize their utilities, based on the current circumstance (e.g., the current $n_l$ and ${\gamma}_{b, s}$). If ${\mathcal{U}}^{b}\le 0$, the trading fails.

\noindent
\textbf{$\bullet$ Futures-based trading without risk evaluation (abbreviate to ``FuturesT-noRE'')}: The two participants negotiate for a forward contract aiming to maximize their expected utilities, without estimating the possible risks (e.g., ${\mathcal{R}}^{b}$, ${\mathcal{R}}^{s}$).

In this simulation, several significant indicators for evaluation and comparison are considered below:

\noindent
\textbf{$\bullet$ Trading Failures (TFail)}: this factor denotes the number of trading failures, which may lead to the unsteadiness of the market.

\noindent
\textbf{$\bullet$ Average buyer attrition rate (ABAR(\%))}: this factor indicates the probability of losing a buyer caused by trading failure, which is defined as the average number of failures per 100 trading.

\noindent
\textbf{$\bullet$  Negotiation latency (NL) and negotiation cost (NC)}: in this simulation, NL is reflected by the running time (millisecond) of the negotiation algorithm for each trading; NC is represented by the number of negotiations per trading (e.g., the value of $i$ in Algorithm 1), which describes the interactive cost (e.g., the possible energy and battery consumption~\cite{27}, and signaling overheads) during negotiation. Apparently, the higher values of NL and NC will lead unsatisfactory user experience to the moving vehicle, which pose challenges to reach a consensus on trading.

\noindent
\textbf{$\bullet$ Trading fairness (TFair)}: in this paper, TFair is defined by the variance of prices $\text{TFair}=\Cov (\mathcal{P})$, which reflects the price fluctuation \cite{16,21} during multiple trading. A smaller variance enables a fairer trading market.

\begin{figure*}[h!t]
\centering
\subfigure[]{\includegraphics[width=.245\linewidth]{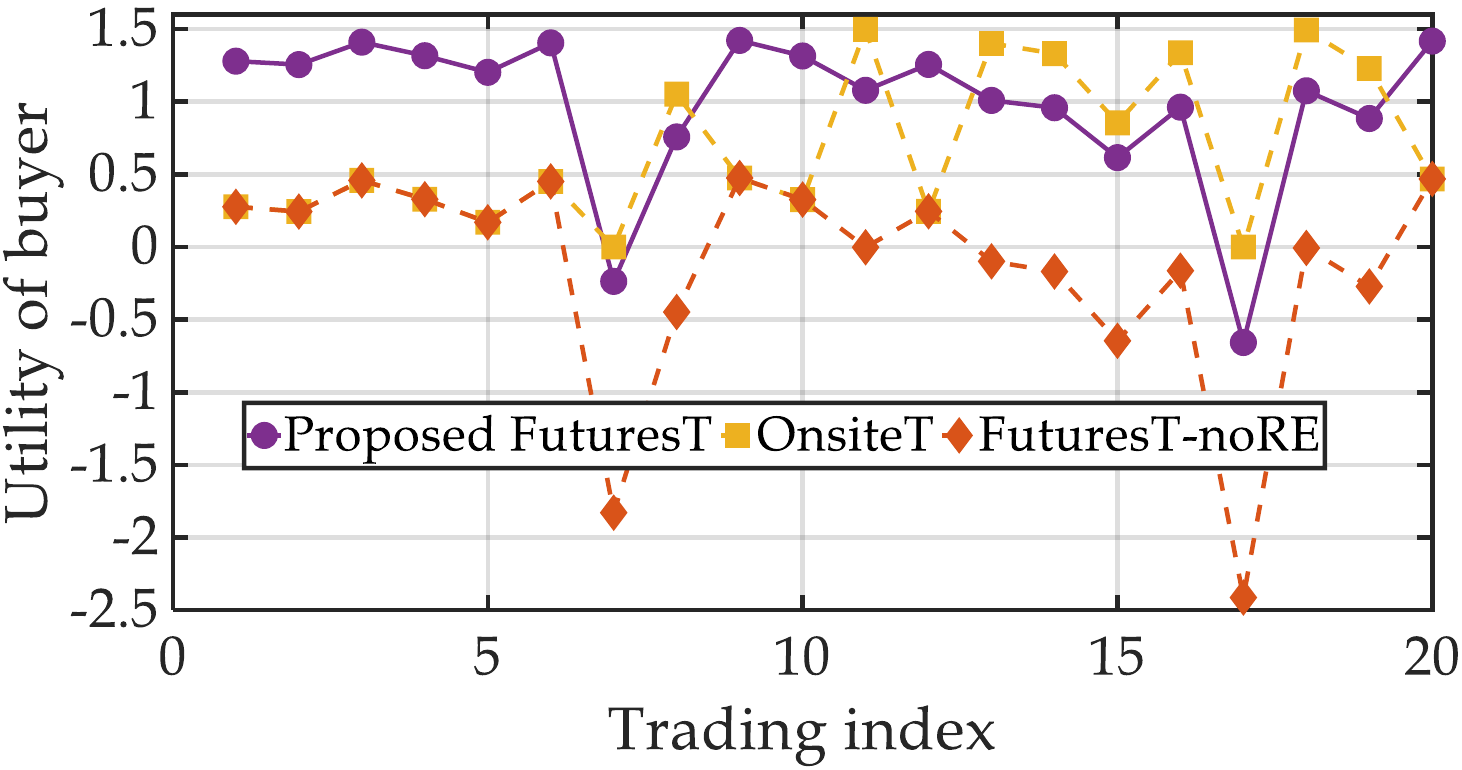}}\hfill
\subfigure[]{\includegraphics[width=.245\linewidth]{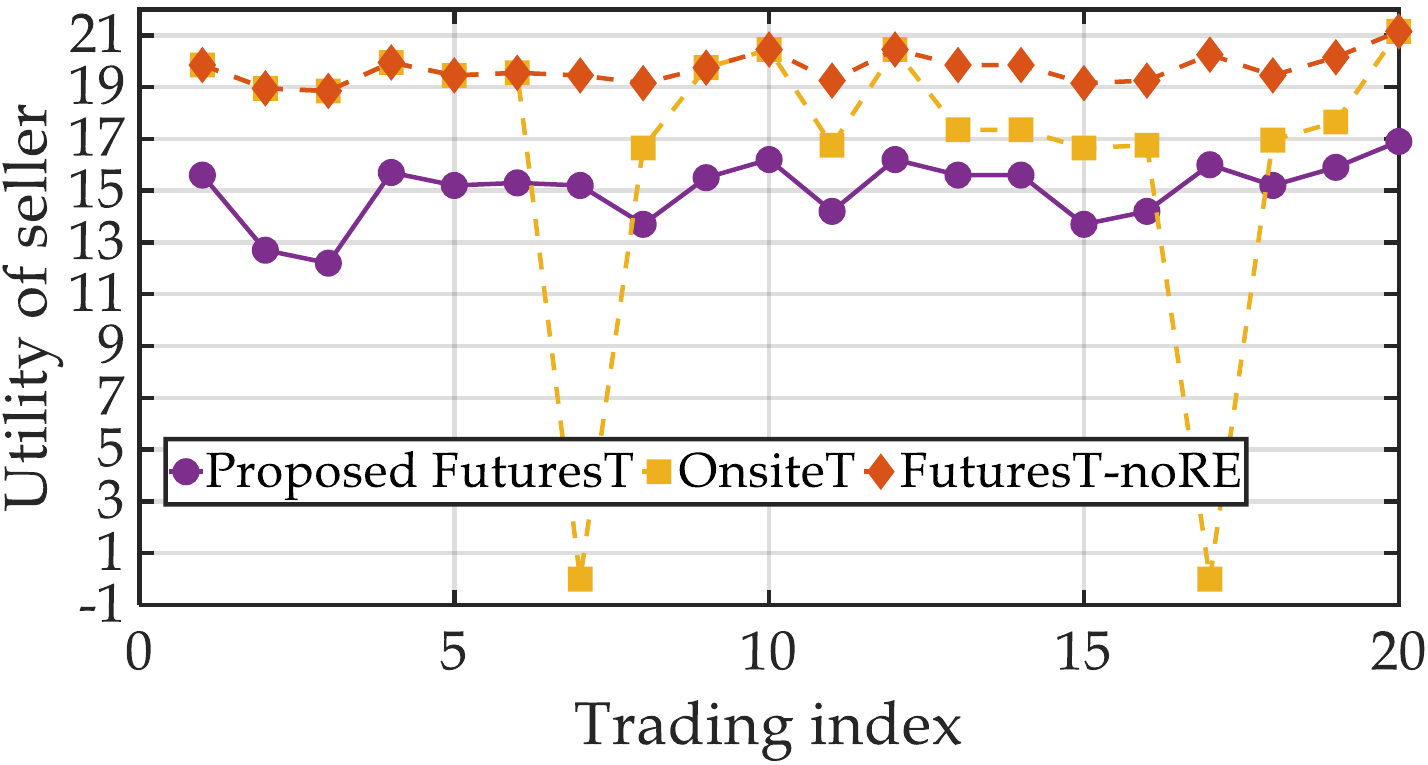}}\hfill
\subfigure[]{\includegraphics[width=.252\linewidth]{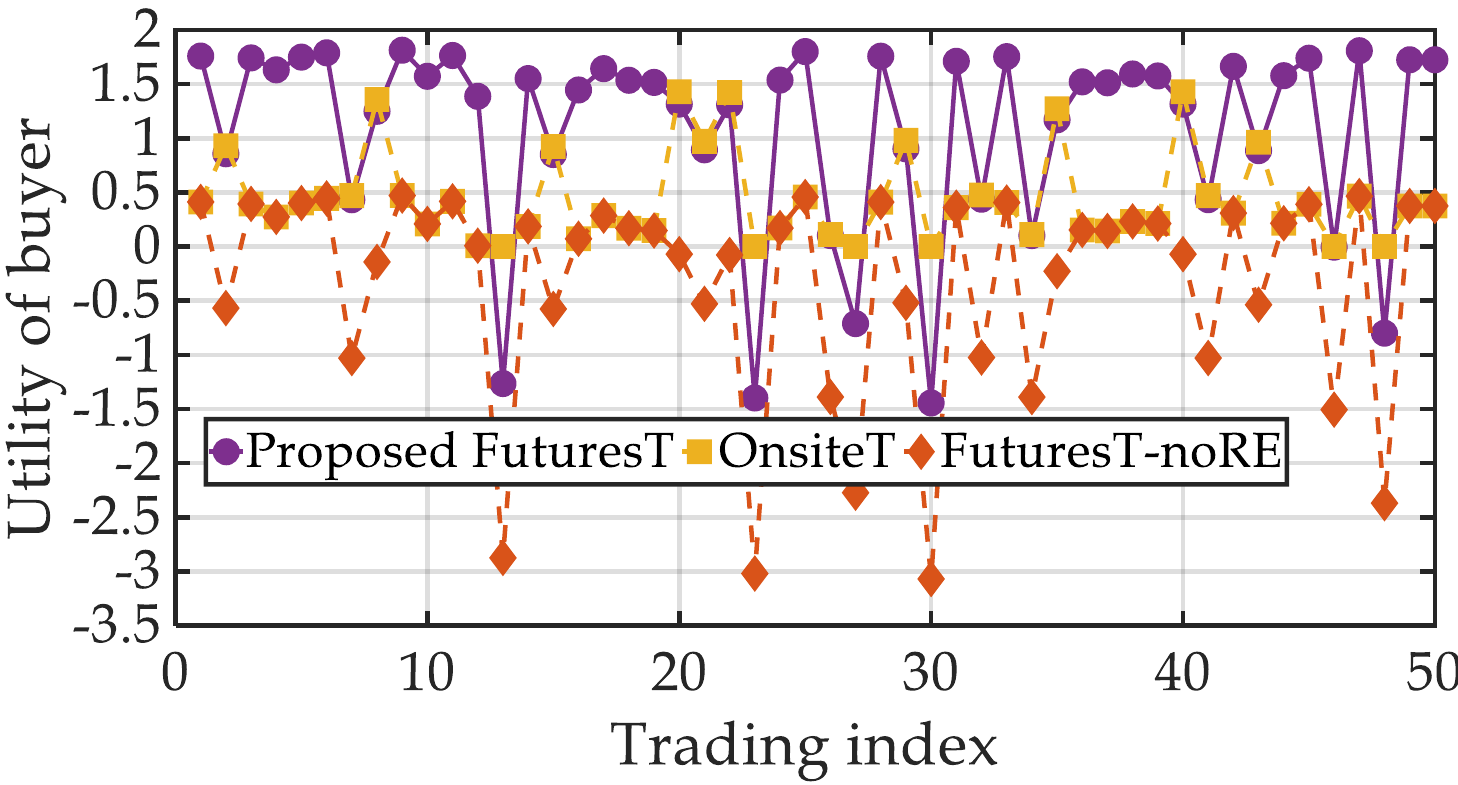}}\hfill
\subfigure[]{\includegraphics[width=.245\linewidth]{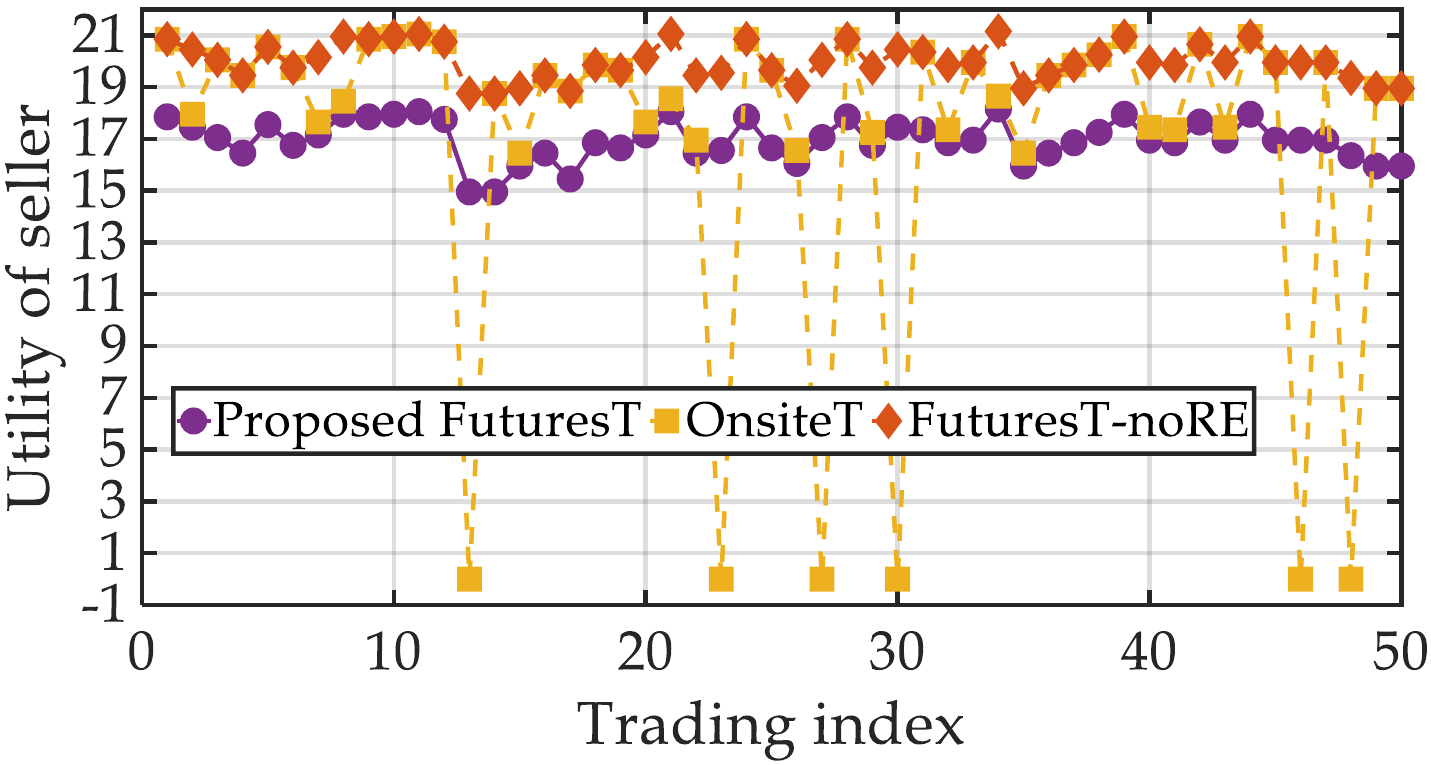}}

\subfigure[]{\includegraphics[width=.255\linewidth]{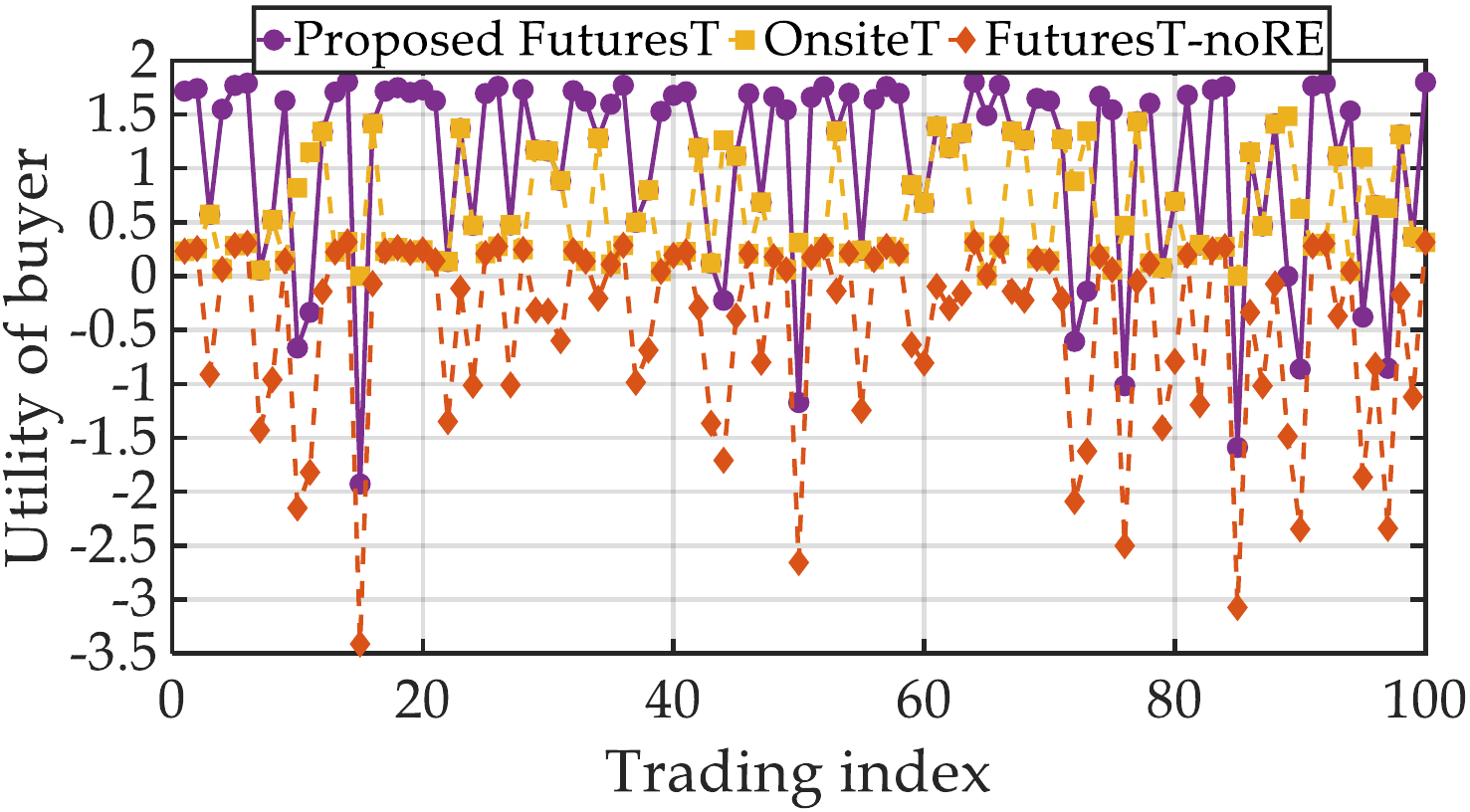}}\hfill
\subfigure[]{\includegraphics[width=.245\linewidth]{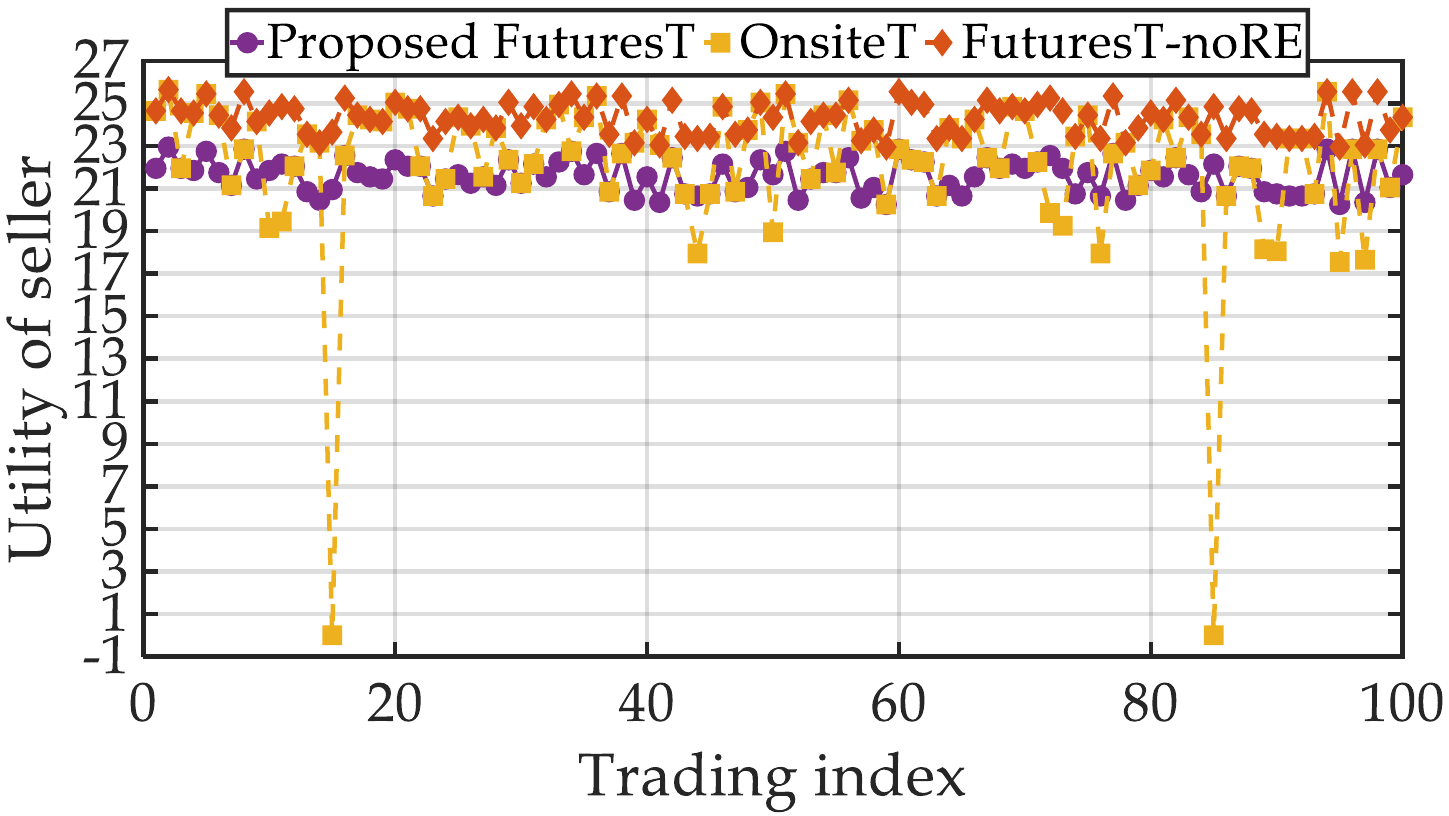}}\hfill
\subfigure[]{\includegraphics[width=.25\linewidth]{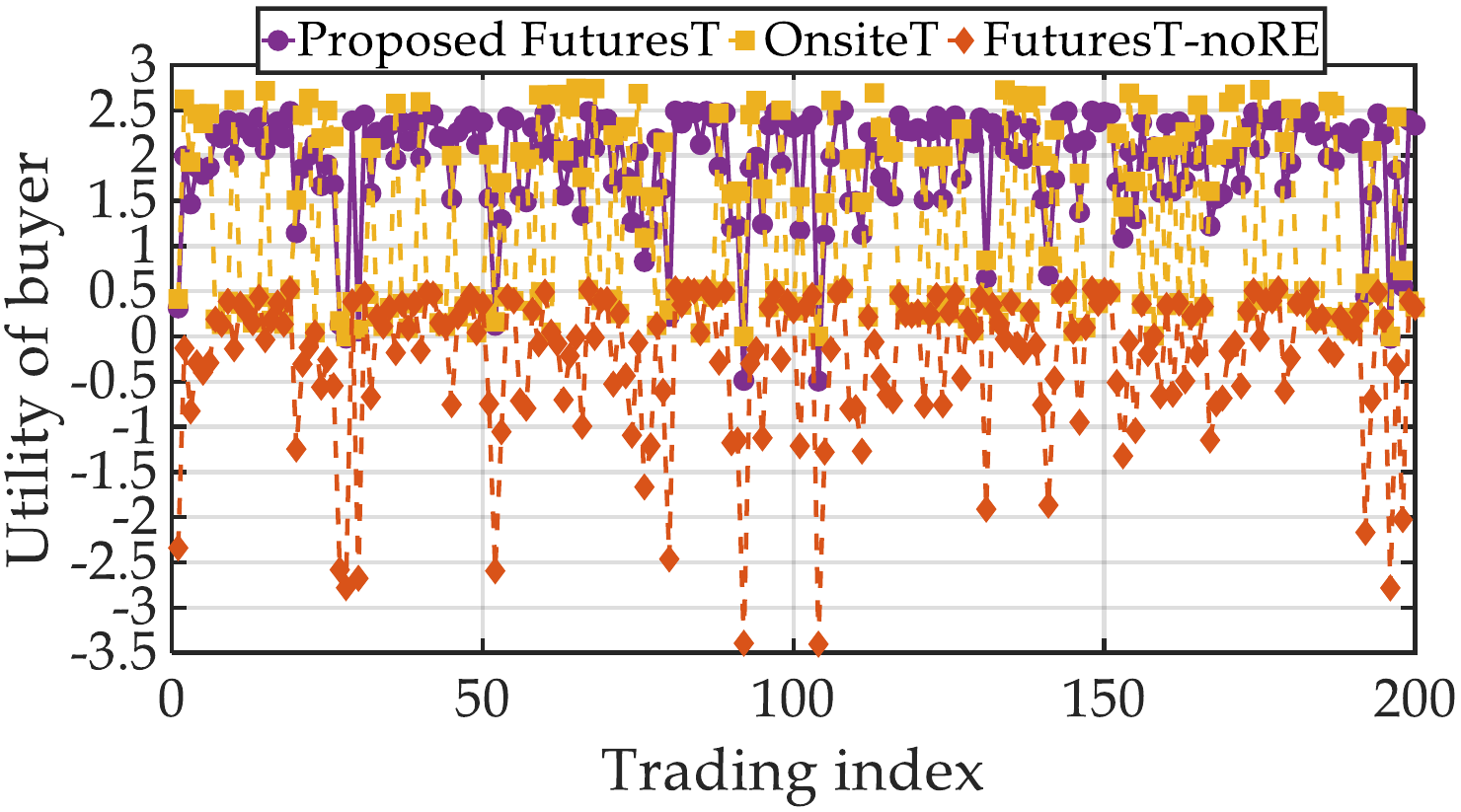}}\hfill
\subfigure[]{\includegraphics[width=.245\linewidth]{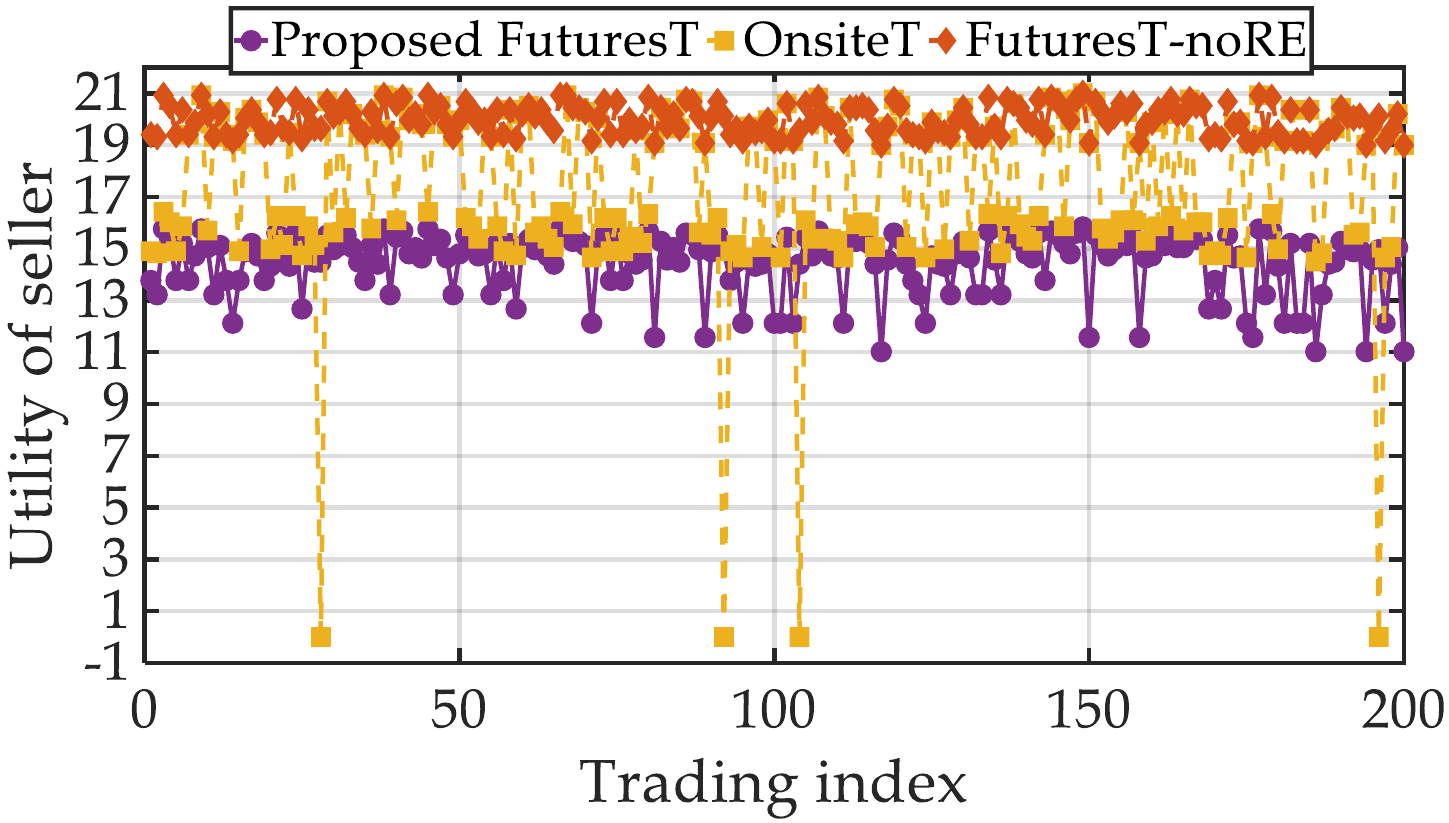}}

\subfigure[]{\includegraphics[width=.245\linewidth]{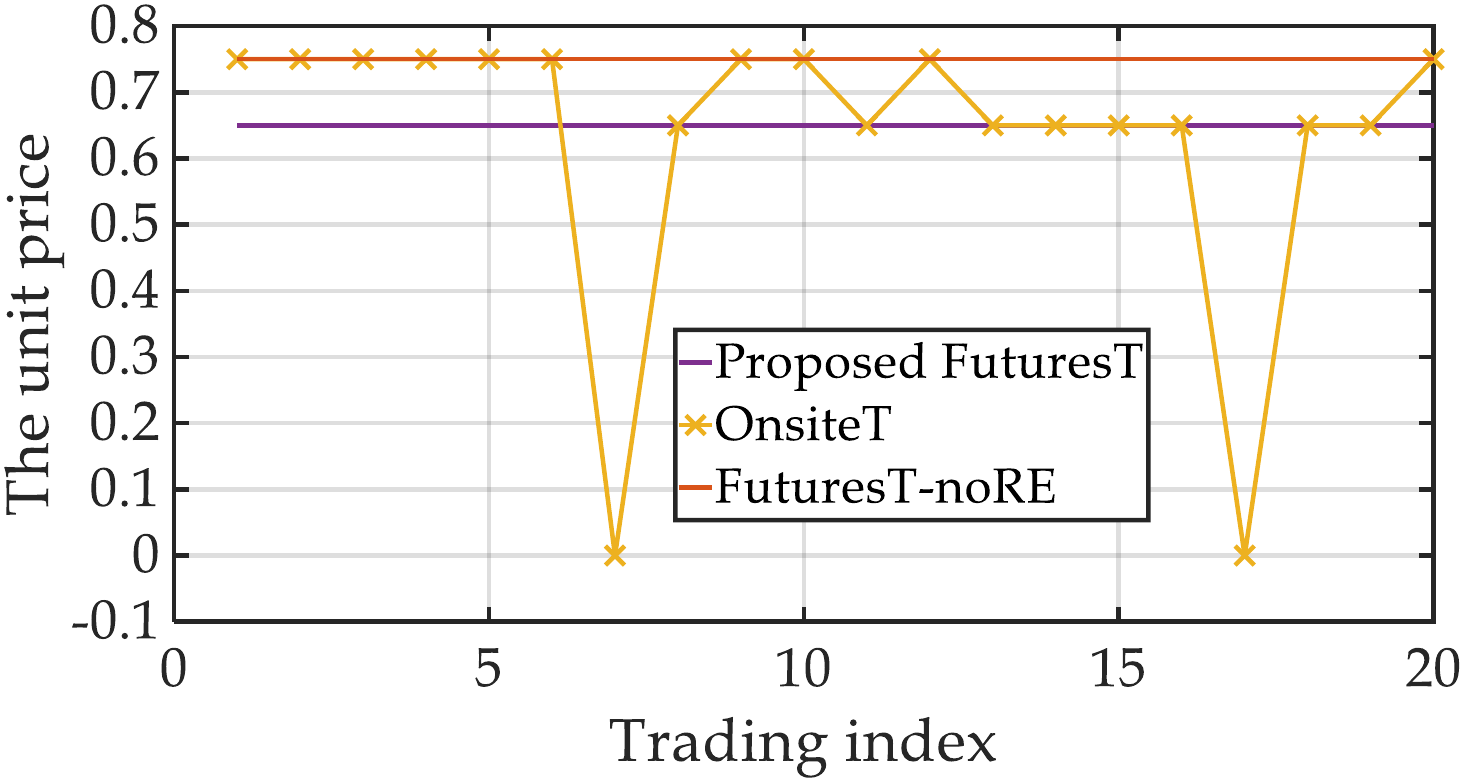}}\hfill
\subfigure[]{\includegraphics[width=.245\linewidth]{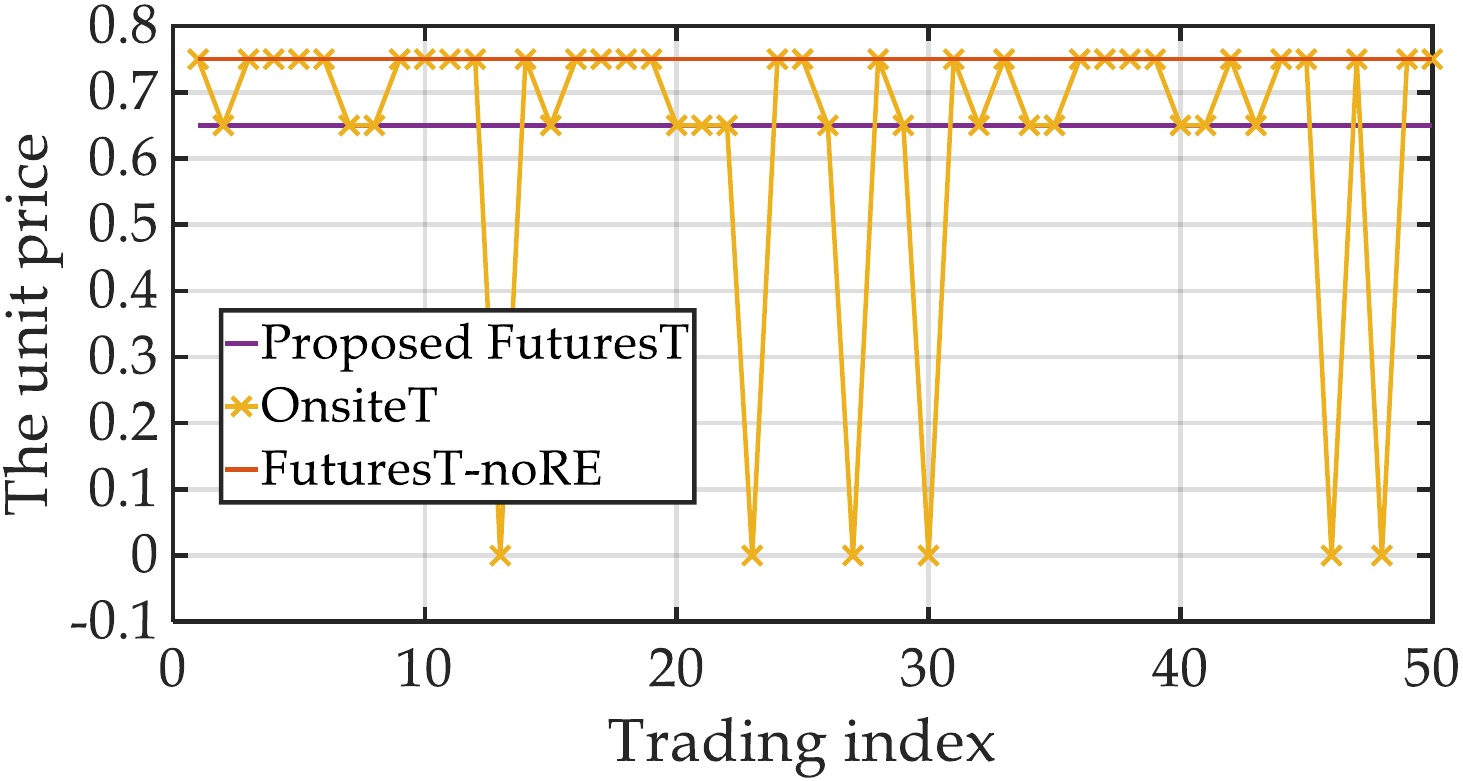}}\hfill
\subfigure[]{\includegraphics[width=.245\linewidth]{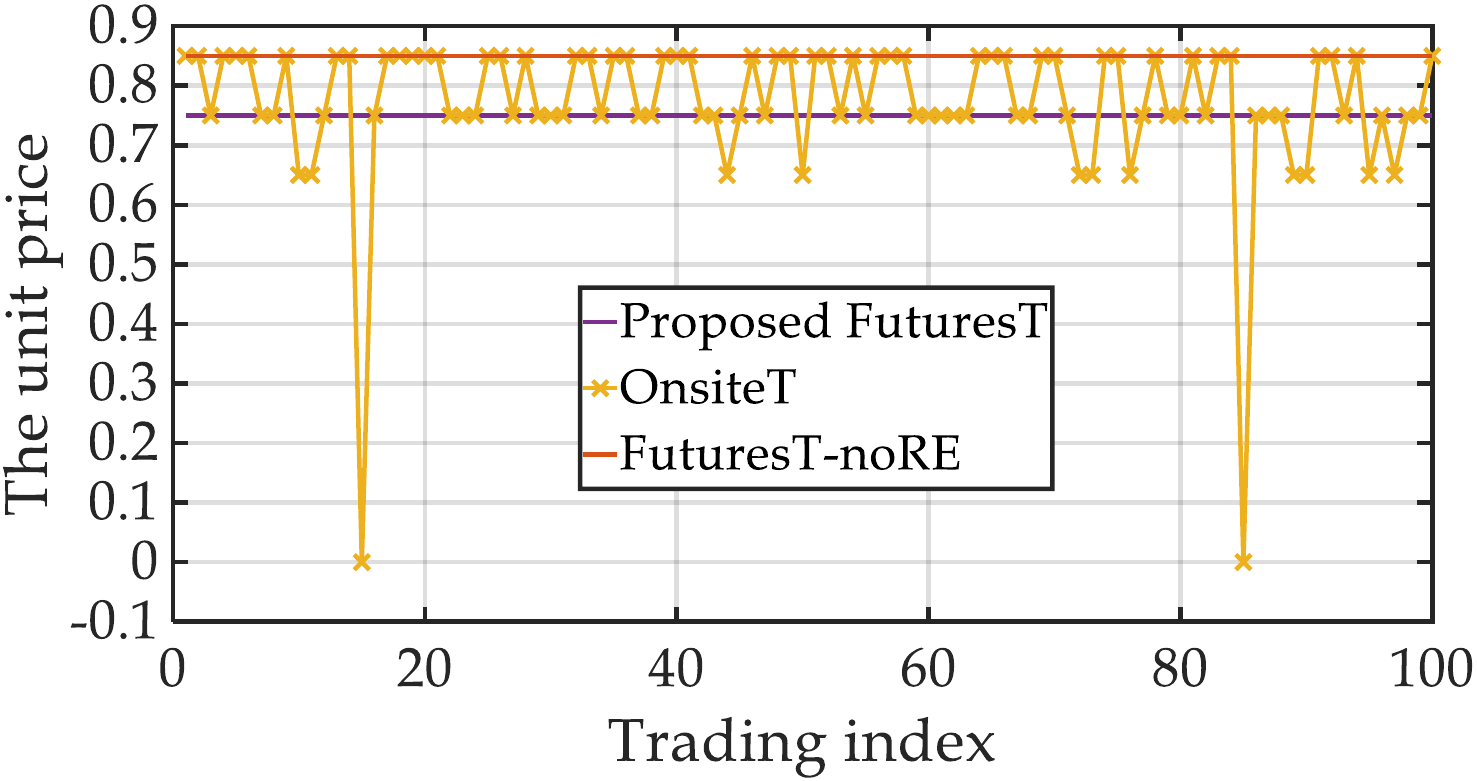}}\hfill
\subfigure[]{\includegraphics[width=.245\linewidth]{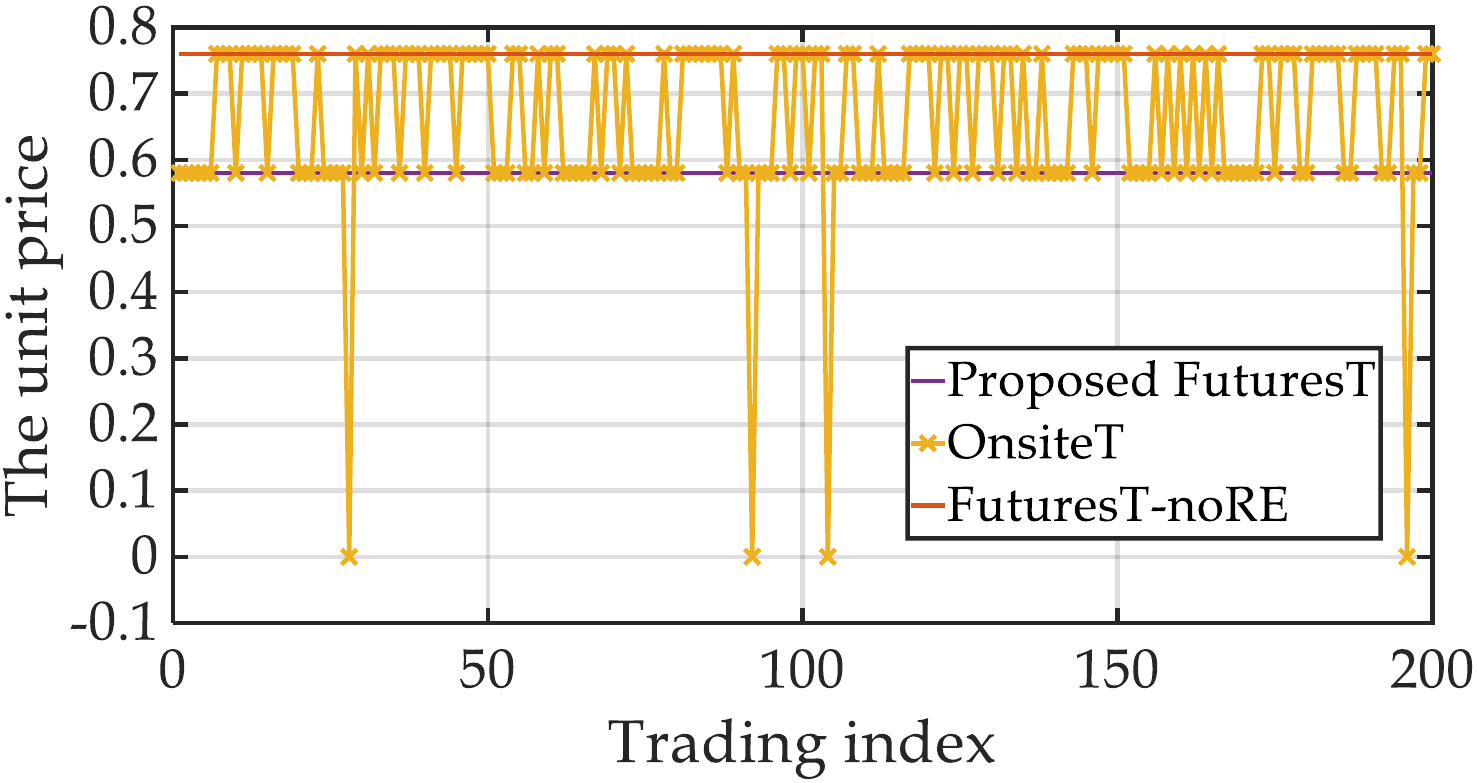}}
\caption{Performance of participants' utilities and the price fluctuation considering different numbers of trading ($M\in \{25,26,27,28,29,30\}$, $\Delta p\in [0.09,0.1]$).}
\end{figure*}


\begin{table*}[h!]
\centering
\caption{Evaluation indicators associated with {\color{black}Fig.}~4 (Algo1: Proposed FuturesT, Algo2: OnsiteT, Algo3: FuturesT-noRE)}
\setlength{\tabcolsep}{1.13mm}{
\begin{tabular}{|*{13}{c|}}
\hline
\textbf{Number of trading} & \multicolumn{3}{c|}{\textbf{20 trading}} & \multicolumn{3}{c|}{\textbf{50 trading}} & \multicolumn{3}{c|}{\textbf{100 trading}} & \multicolumn{3}{c|}{\textbf{200 trading}} \\
\hline
\textbf{Methods} & Algo1 & Algo2 & Algo3 & Algo1 & Algo2 & Algo3 & Algo1 & Algo2 & Algo3 & Algo1 & Algo2 &Algo3  \\ \hline
\textbf{TFail} & 0 & 2 & 0 & 0 & 6 & 0 & 0 & 2 & 0 & 0 & 4 & 0 \\ \hline
\textbf{ABAR}  & 0 & 10\% & 0 & 0 & 12\% & 0 & 0 & 2\% & 0 & 0 & 2\% & 0 \\ \hline
\textbf{NC} & 3 & 28 & 3 & 3 & 73 & 3 & 3 & 232 & 2 & 2 & 299 & 2 \\ \hline
\textbf{TFair} $\left(\times 10^{-2}\right)$ & 0 & 4.95 & 0 & 0 & 5.9 & 0 & 0 & 1.68 & 0 & 0 & 1.69 & 0 \\ \hline
\textbf{Sum} $\bm{({\mathcal{U}}^{b})}$ & 19.72 & 13.66 & $-$2.61 & 54.54 & 21.90 & $-$15.72 & 104.72 & 54.78 & $-$43.57 & 384.58 & 223.71 & $-$44.76 \\ \hline
\textbf{Sum} $\bm{({\mathcal{U}}^{s})}$ & 300.80 & 344.51 & 394.28 & 848.10 & 844.50 & 1000.10 & 2160.71 & 2214.85 & 2430.71 & 2890.36 & 3495.19 & 3993.03 \\
\hline
\end{tabular}}
\label{tab2}
\end{table*}

\subsection{Performance comparison and evaluation}

\begin{figure*}[h!t]
\centering
\subfigure[]{\includegraphics[width=.245\linewidth]{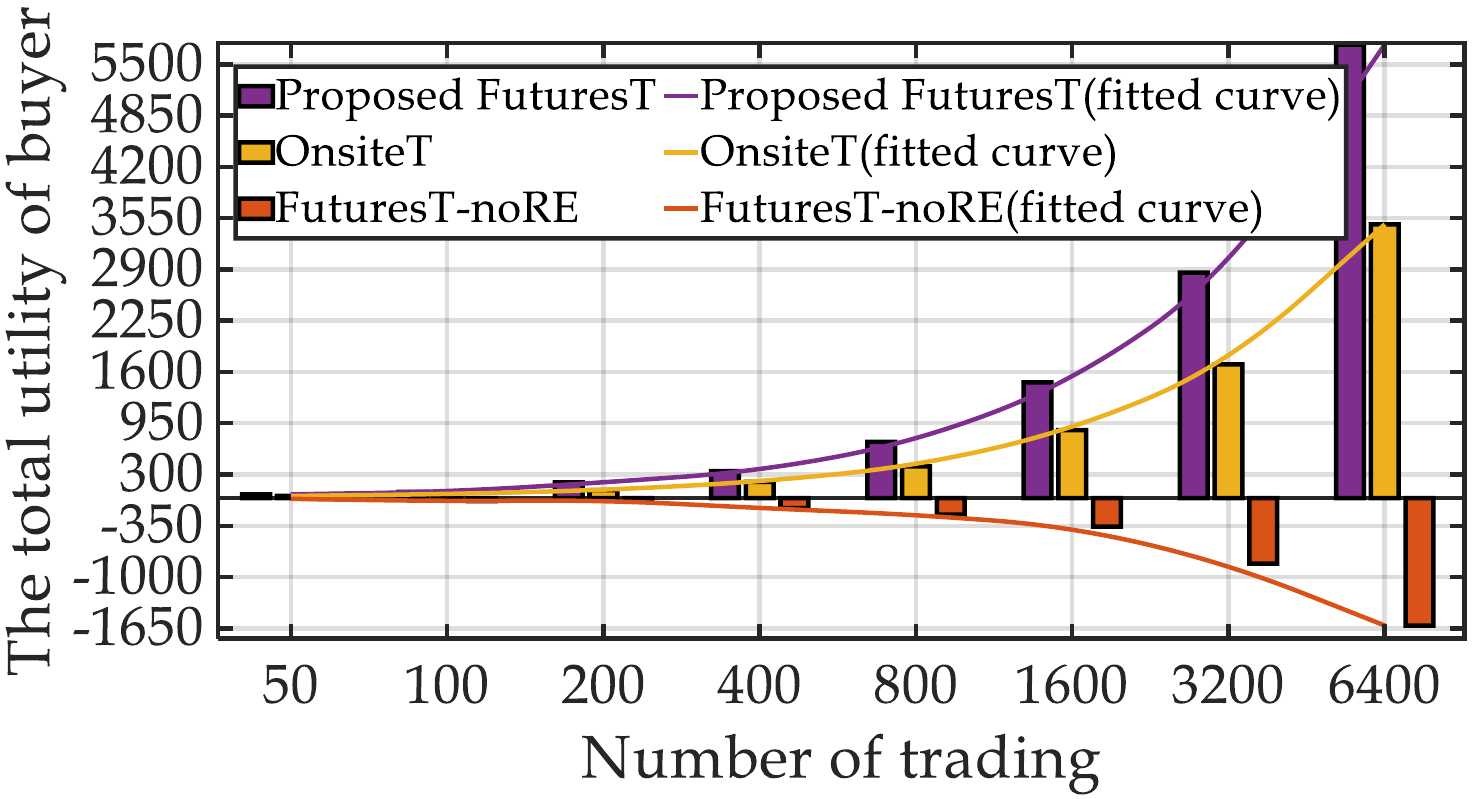}}\hfill
\subfigure[]{\includegraphics[width=.255\linewidth]{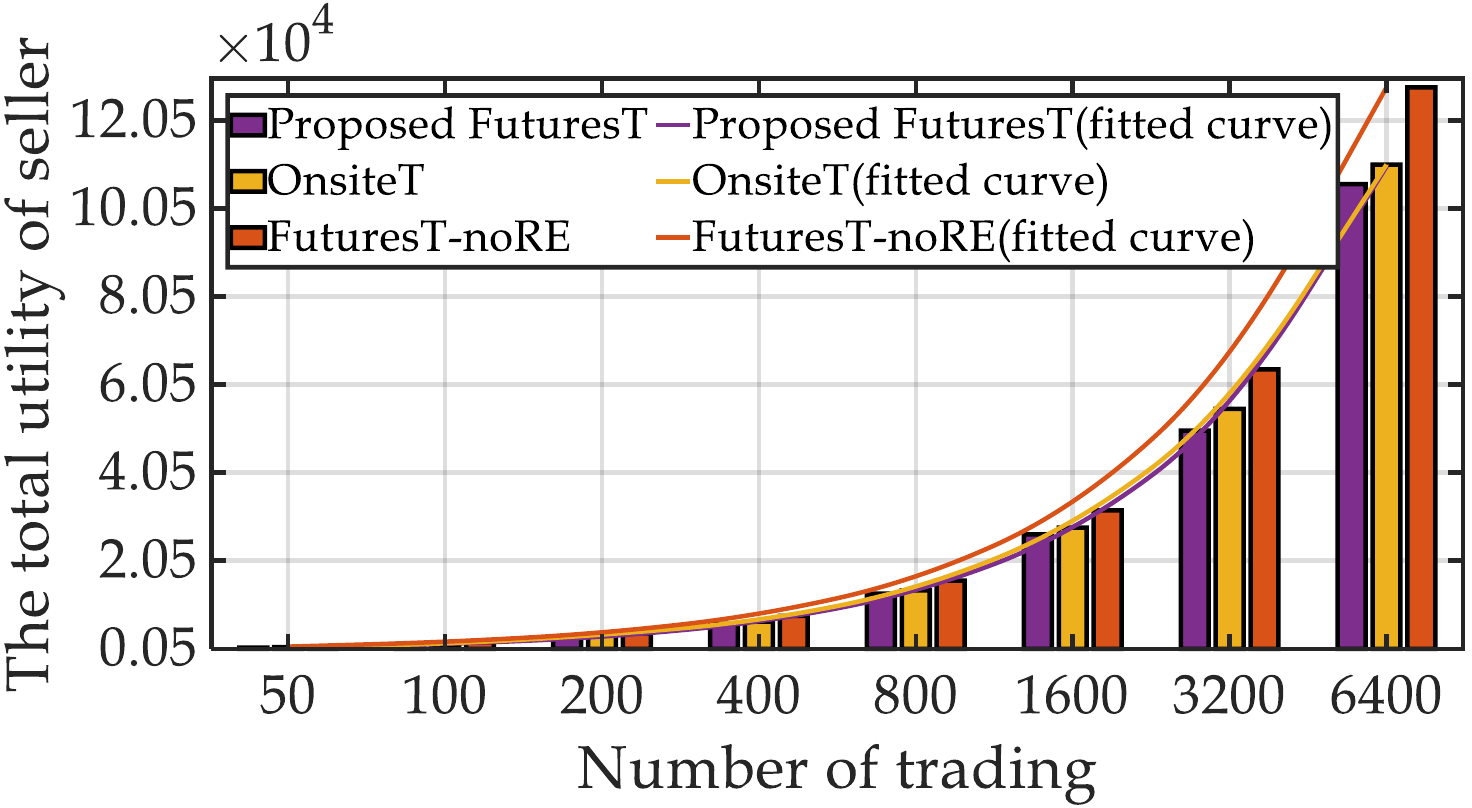}}\hfill
\subfigure[]{\includegraphics[width=.243\linewidth]{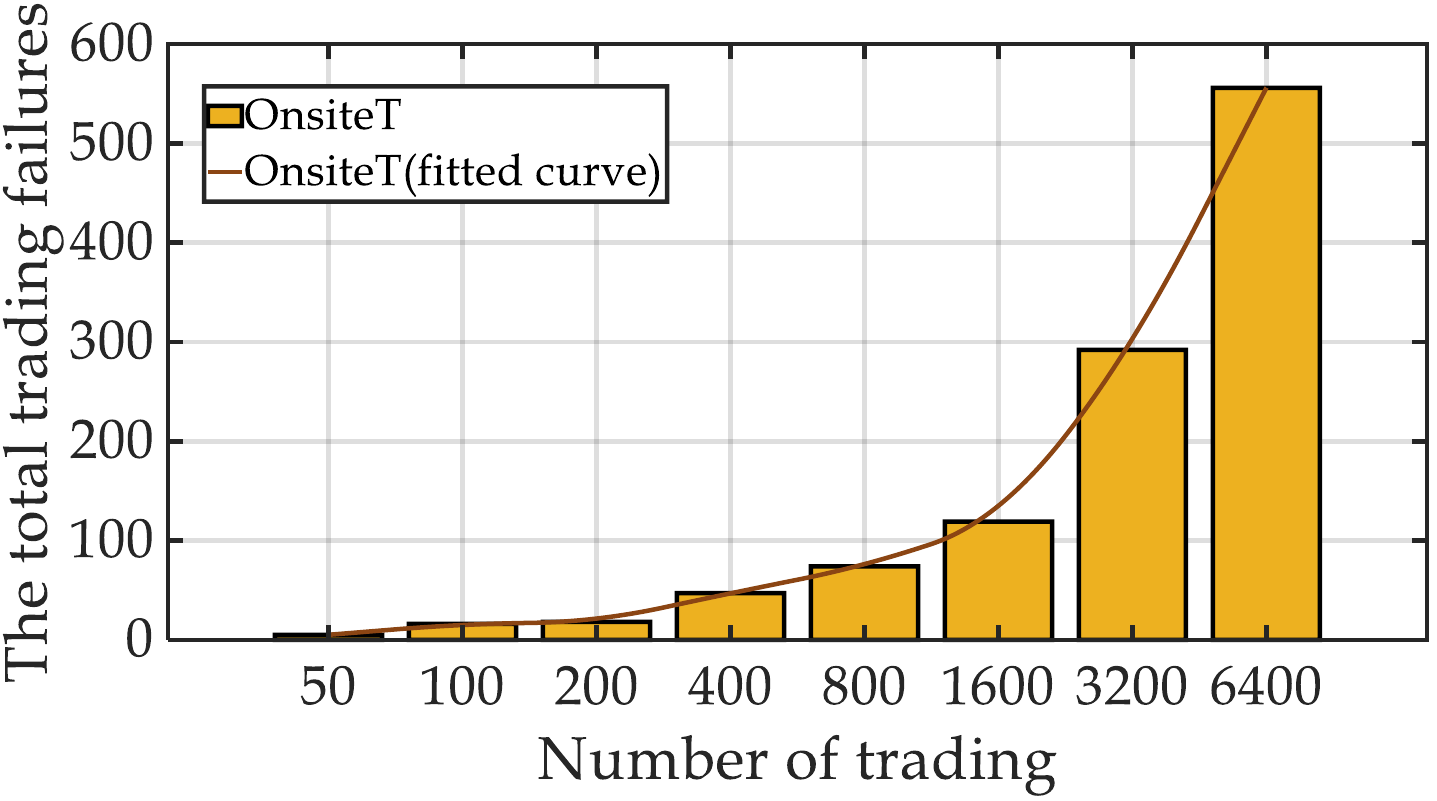}}\hfill
\subfigure[]{\includegraphics[width=.25\linewidth]{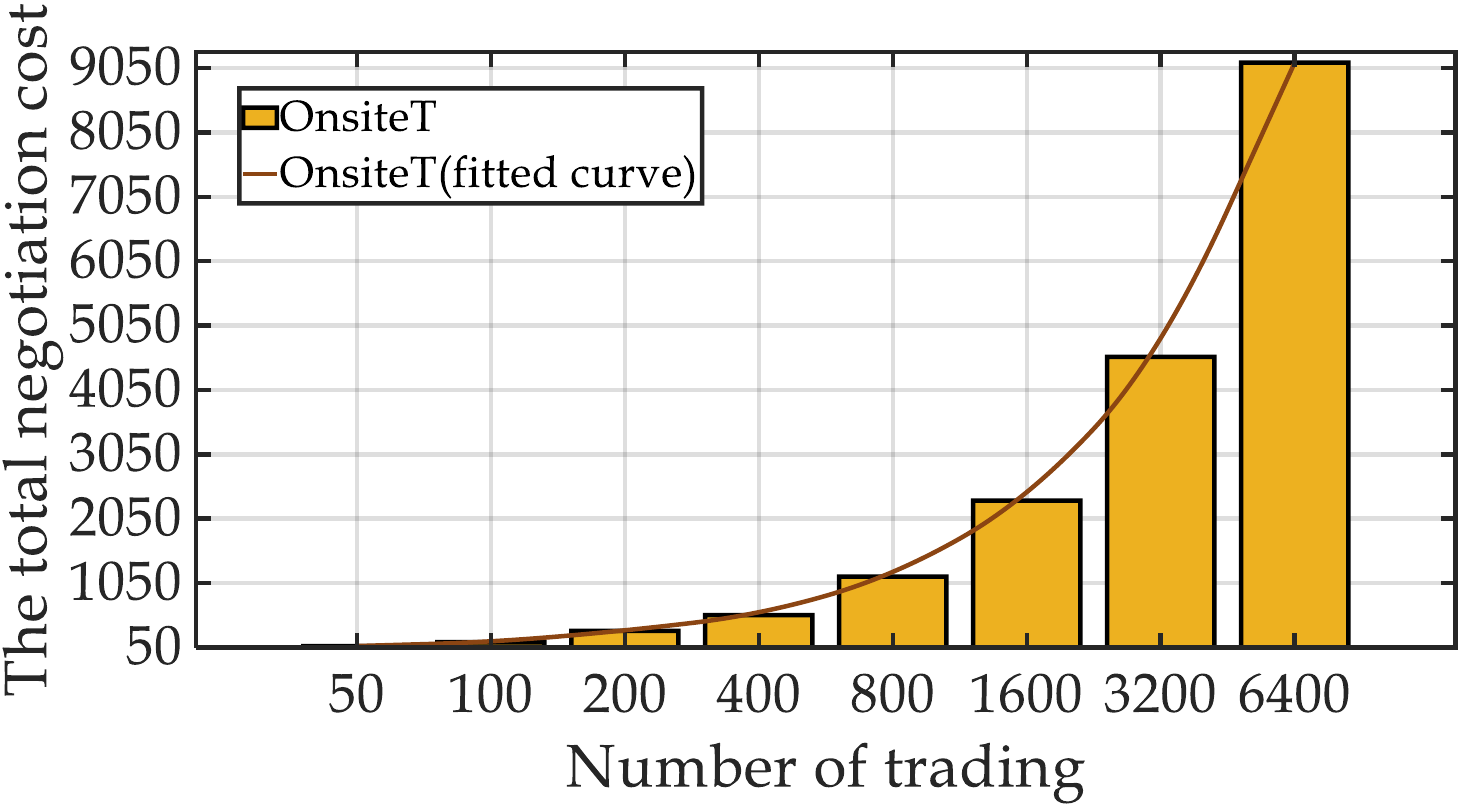}}
\caption{Performance evaluation of long-term utilities, failures, and negotiation costs ($M\in \{25,26,27,28,29,30\}$, $\Delta p\in [0.1,0.2]$).}

\end{figure*}
\begin{figure*}[h!t]
\centering
\subfigure[]{\includegraphics[width=.249\linewidth]{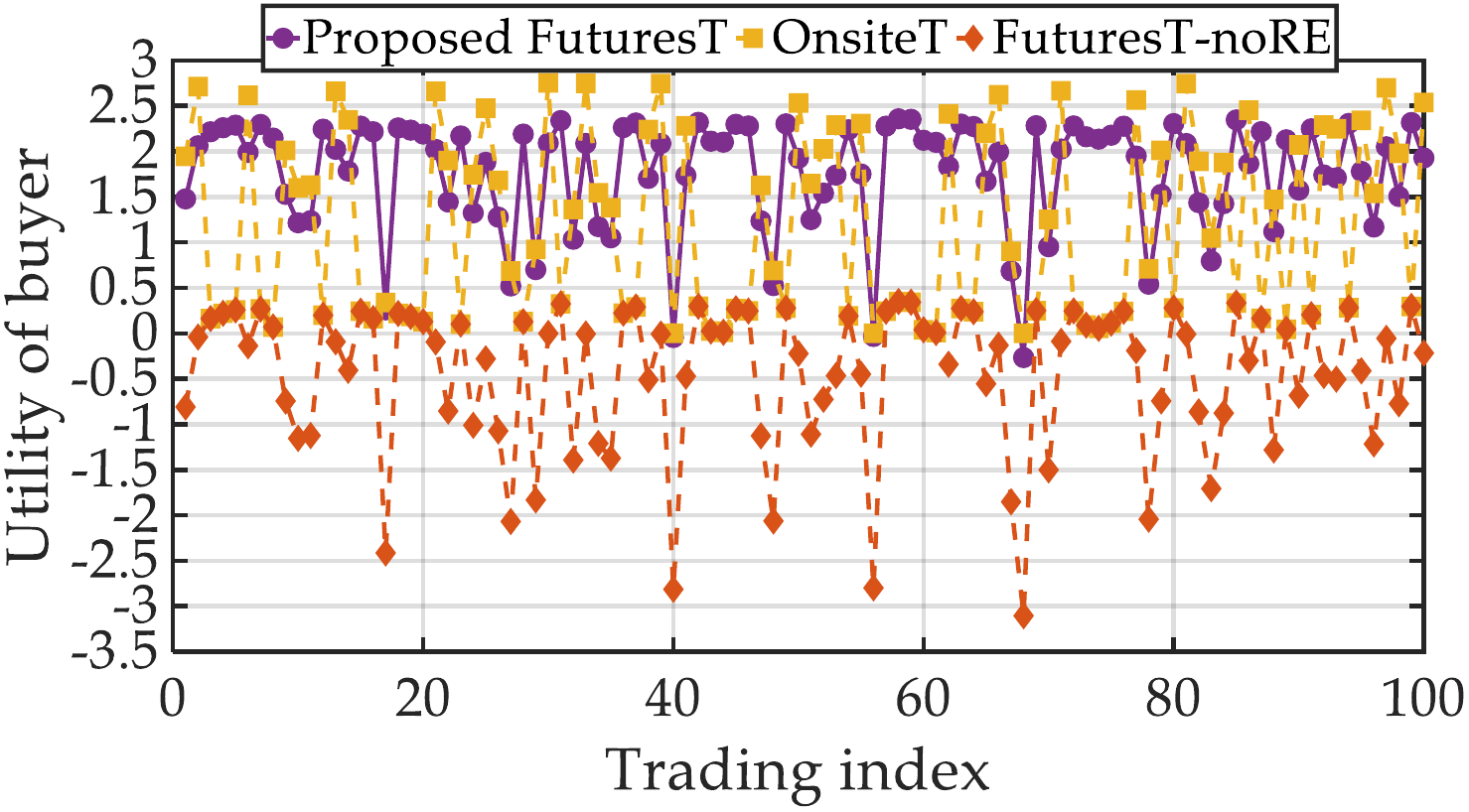}}\hfill
\subfigure[]{\includegraphics[width=.245\linewidth]{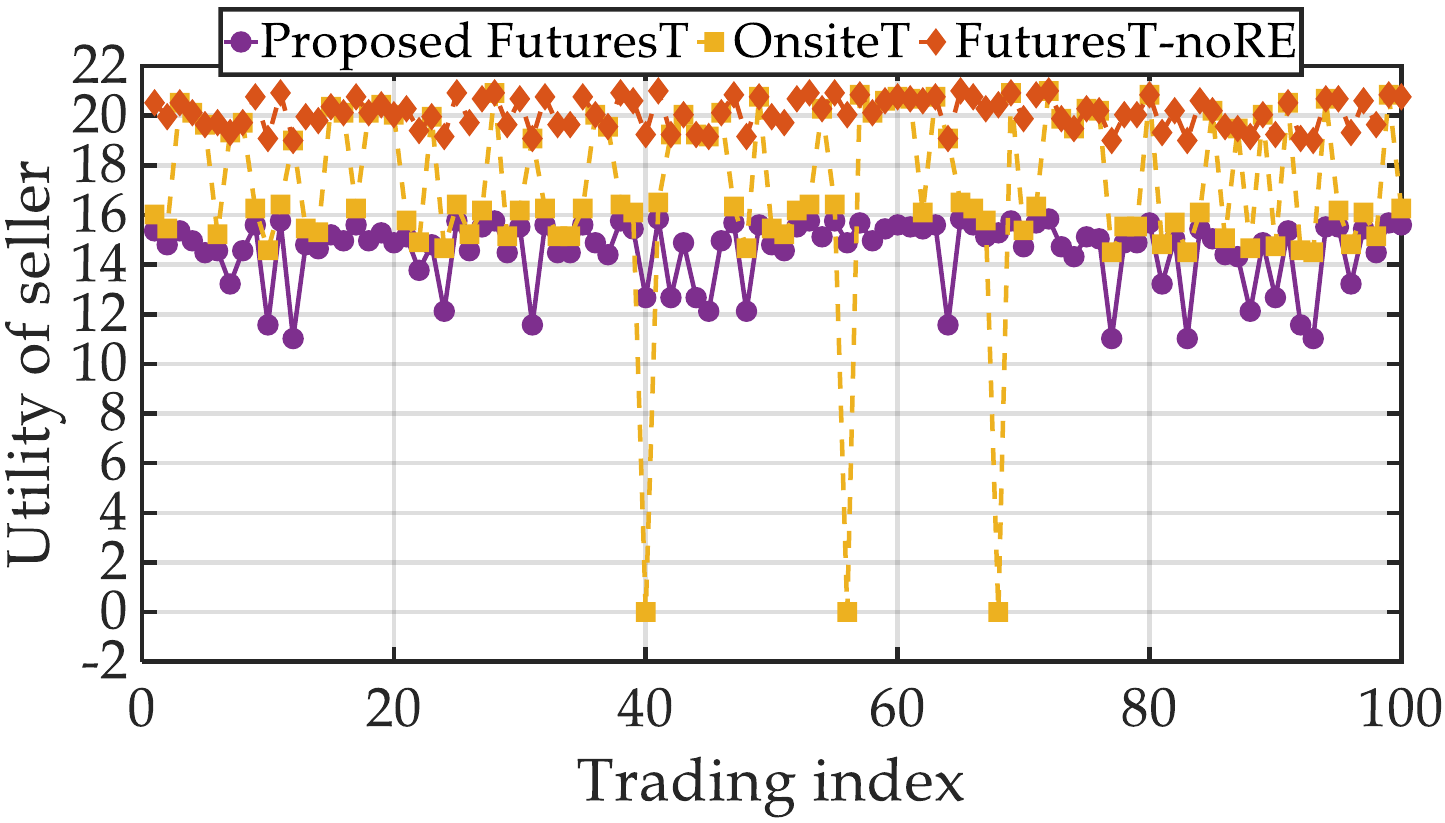}}\hfill
\subfigure[]{\includegraphics[width=.25\linewidth]{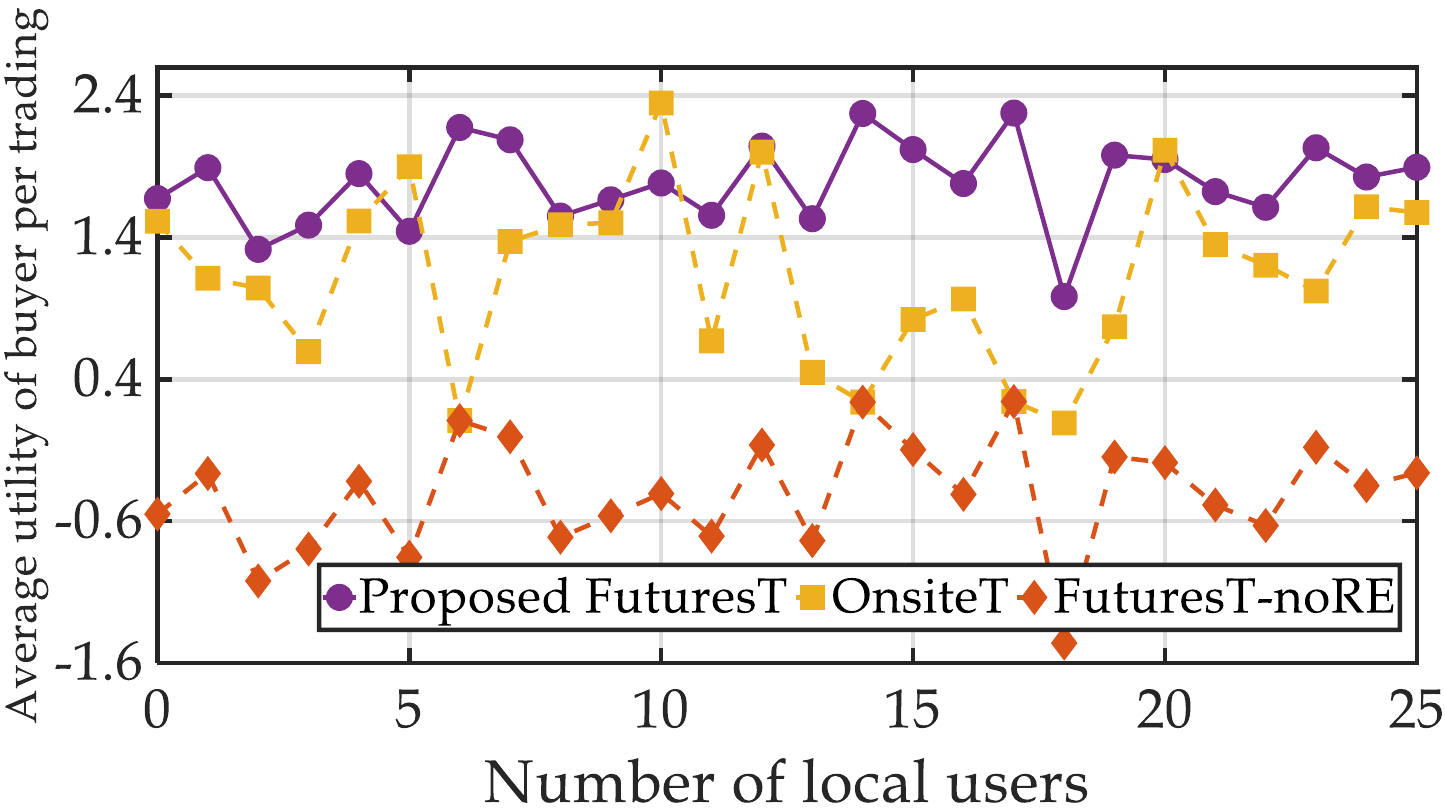}}\hfill
\subfigure[]{\includegraphics[width=.245\linewidth]{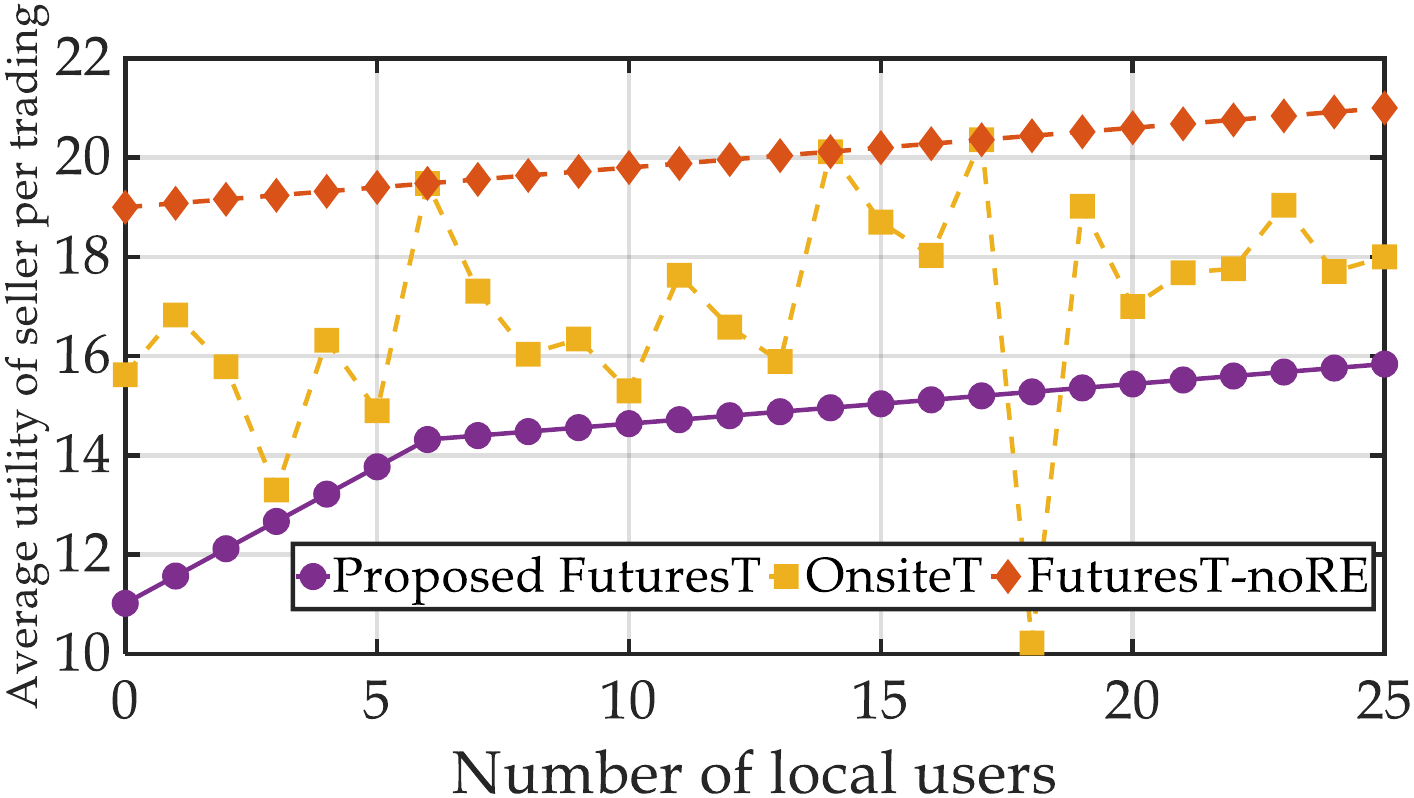}}

\subfigure[]{\includegraphics[width=.245\linewidth]{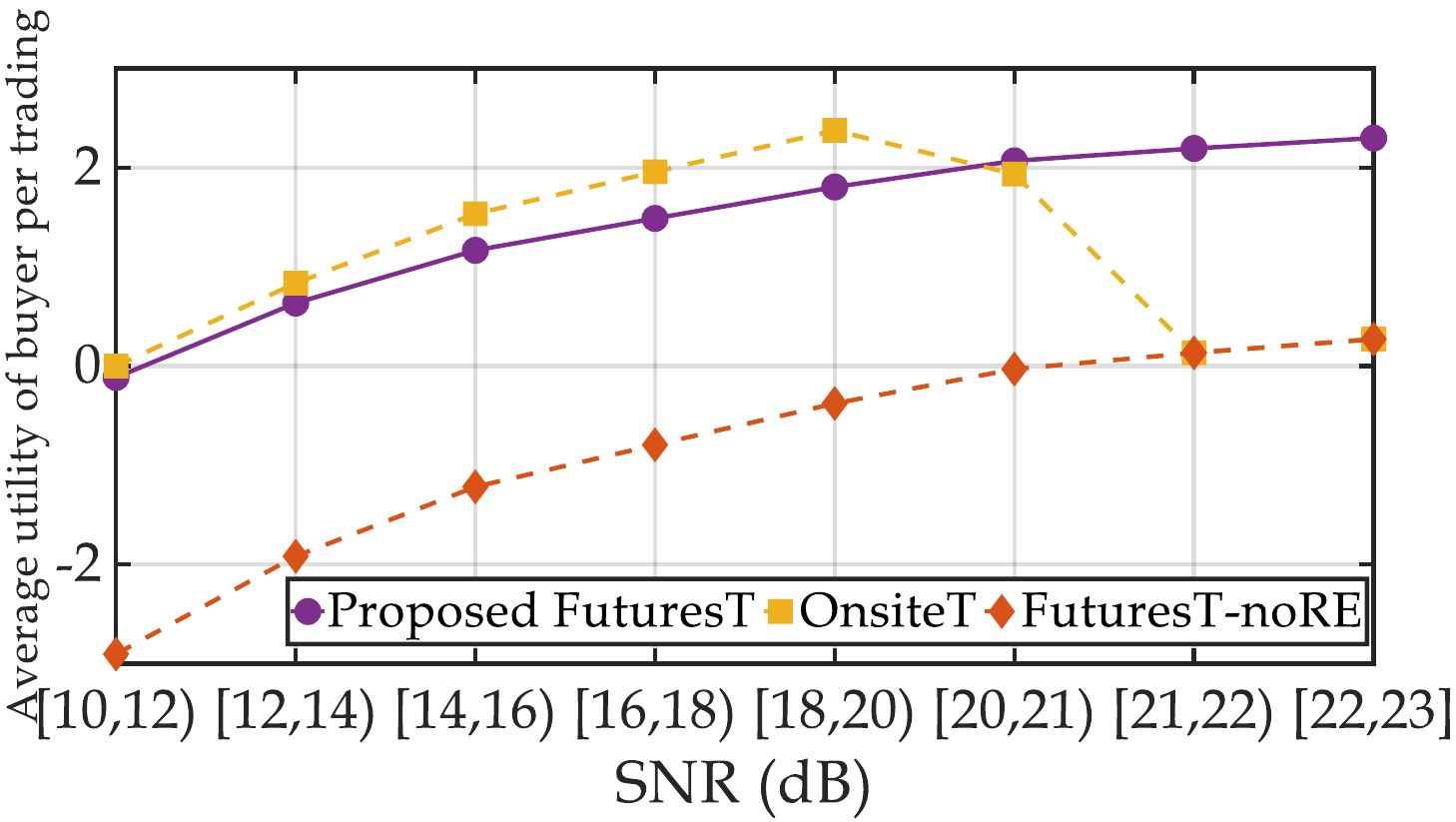}}\hfill
\subfigure[]{\includegraphics[width=.247\linewidth]{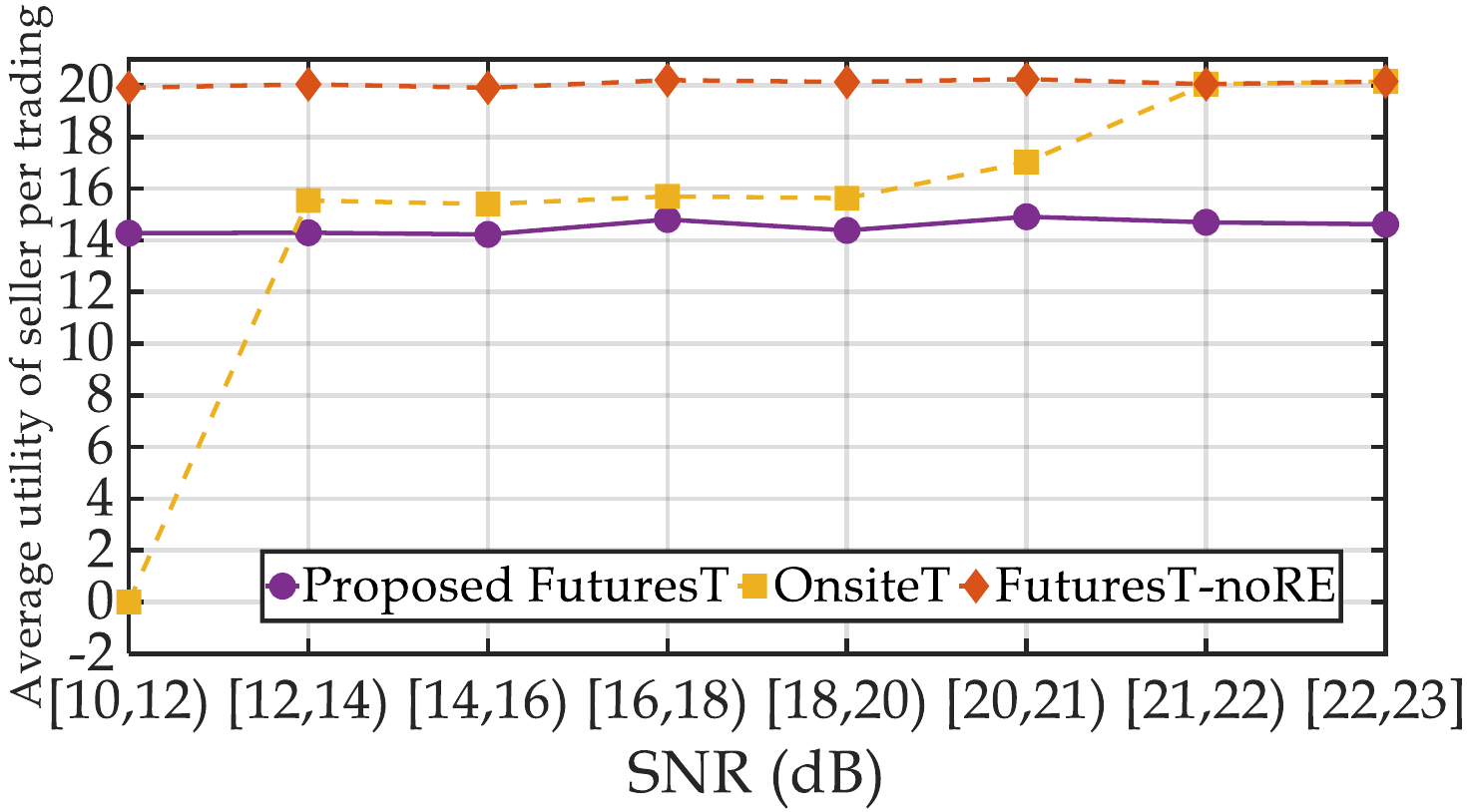}}\hfill
\subfigure[]{\includegraphics[width=.249\linewidth]{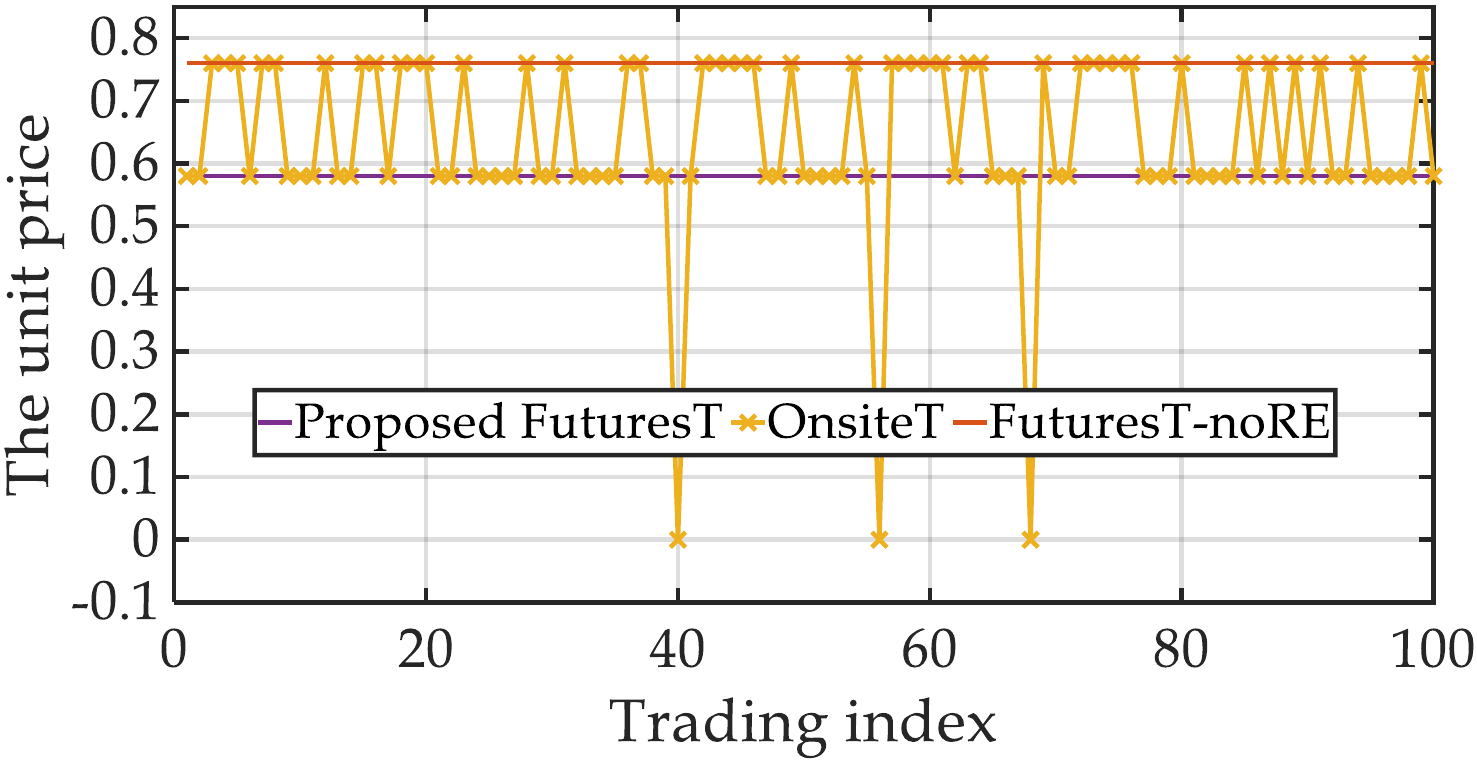}}\hfill
\subfigure[]{\includegraphics[width=.249\linewidth]{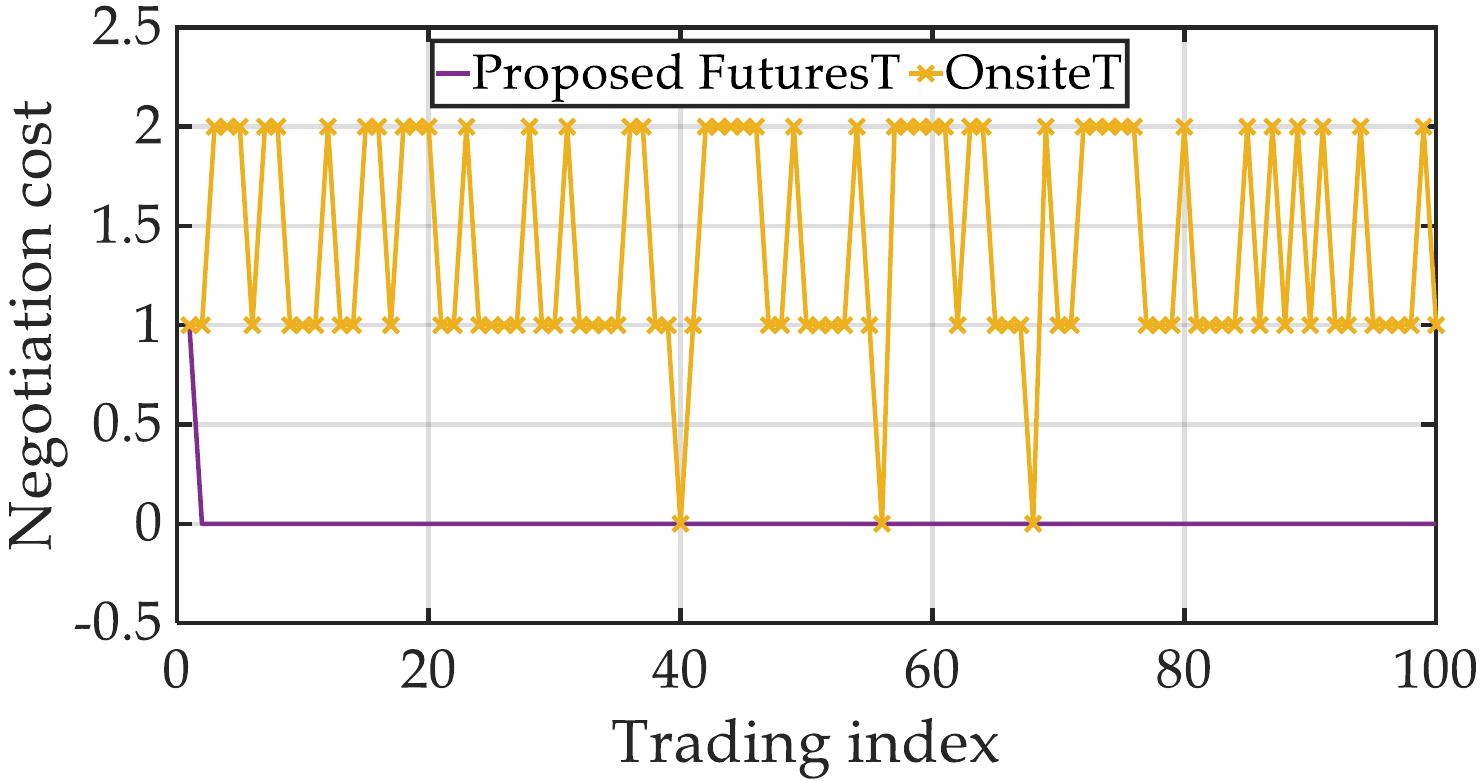}}

\caption{Performance comparisons upon having various $n_l$ and ${\gamma}_{b, s}$ ($M=25$, $\Delta p\in [0.09,0.18]$).}
\end{figure*}
Figure~4 shows the performance of the proposed futures-based resource trading approach on participants' utilities and price fluctuation, comparing with the baseline methods under different numbers of trading: 20 trading (see Fig.~4(a), Fig.~4(b) and Fig.~4(i)), 50 trading (see Fig.~4(c), Fig.~4(d) and Fig.~4(j)), 100 trading (see Fig.~4(e), Fig.~4(f) and Fig.~4(k)), 200 trading (see Fig.~4(g), Fig.~4(h) and Fig.~4(l)). 

Specifically, Fig.~4(a), Fig.~4(c), Fig.~4(e) and Fig.~4(g) show that our proposed approach facilitates higher utilities for the buyer in most trading rather than the other methods, although sometimes obtaining negative utilities owing to the prediction uncertainty of $\gamma_{b, s}$. Moreover, the onsite-based method suffers from failures due to that a successful onsite trading can only be fulfilled relying on the current network status. FuturesT-noRE always receives negative values of the buyer's utility owing to the negligence of possible risks. Fig.~4(b), Fig.~4(d), Fig.~4(f) and Fig.~4(h) depict the performance comparisons of the seller's utility considering various numbers of trading. In these figures, the curves of FuturesT-noRE stay above that of the other methods in most trading, due to that a larger $\mathcal{A}$ brings the buyer a larger expected utility under fixed $\mathcal{P}$ (when $\partial {\overline{\mathcal{U}^{b}}}/{\partial \mathcal{A}}>0$), as observed in Fig.~3(d); meanwhile, the seller tends to select a higher price when $\mathcal{A}$ is fixed (see Fig.~3(c)). Accordingly, the forward contract in FuturesT-noRE is more inclined to maximize the seller' utility, which, however, sacrifices the buyer's utility. 

The proposed FuturesT achieves considerable performance of the seller's utility although sometimes inferior to the OnsiteT method, which is caused by the prediction uncertainties of the current network status. Apparently from Fig.~4(b), Fig.~4(d), Fig.~4(f) and Fig.~4(h), the onsite trading often fails owing to the possible factors such as a small $\gamma_{b, s}$ and a large $p^{min}_{s}$, etc. Moreover, the onsite negotiation procedure may lead to larger negotiation cost and latency. Fig.~4(i), Fig.~4(j), Fig.~4(k) and Fig.~4(l) present the price fluctuation under different numbers of trading. Apparently, the prices of the two futures-based approaches remain unchanged, which facilitate a stable resource trading market. However, the prices of OnsiteT keeps fluctuating, and thus will introduce a worse fairness.

%
%
%

Table~2 associated with {\color{black}Fig.}~4 shows the performance of the indicators TFail, ABAR, NC, TFair, and the sum of ${\mathcal{U}}^{b}$ and ${\mathcal{U}}^{s}$. Observing from Table~2,  OnsiteT sustains various number of trading failures, which thus causes higher ABARs, and an unstable resource trading market. The negotiation for the forward contract only happens once between the futures-based participants, and thus can achieve far less NC comparing with the onsite-based method. Associated with Fig.~4, a larger TFair stands for a severer fluctuation of prices, which correspondingly leads worse fairness to the market. As for the sum of participants' utilities, the proposed FuturesT outperforms the other methods on the buyer's utility, and achieves similar performance on the sum of seller's utilities with the OnsiteT method. However, the proposed FuturesT approach obtains far better performance on TFail, ABAR, NC, and TFair than OnsiteT. The FuturesT-noRE method gets higher seller's utility by sacrificing the buyer's, which poses challenges maintaining the stability of the resource trading environment.

\begin{figure*}[h!t]
\centering
\subfigure[]{\includegraphics[width=.345\linewidth]{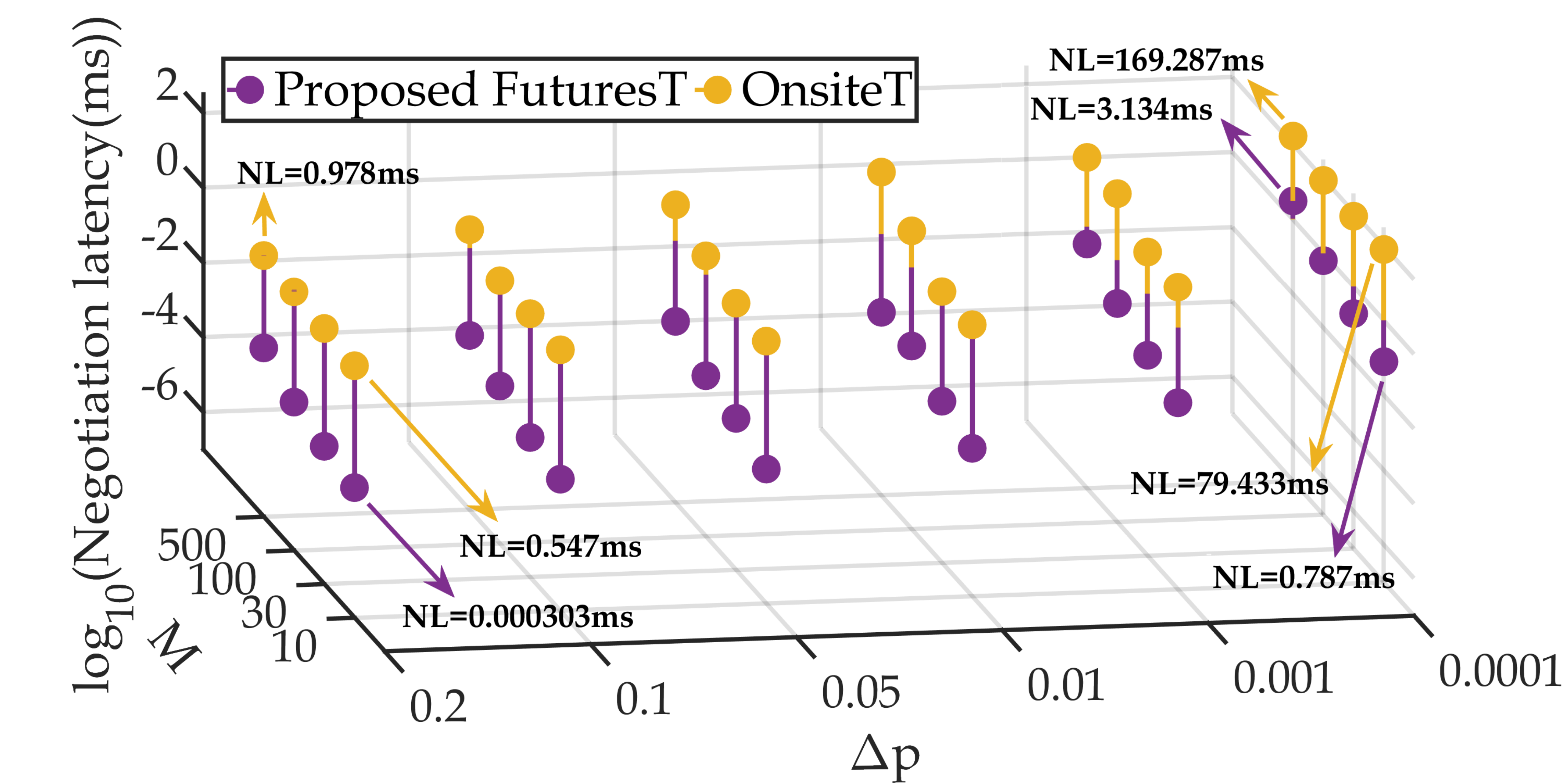}}\hfill
\subfigure[]{\includegraphics[width=.345\linewidth]{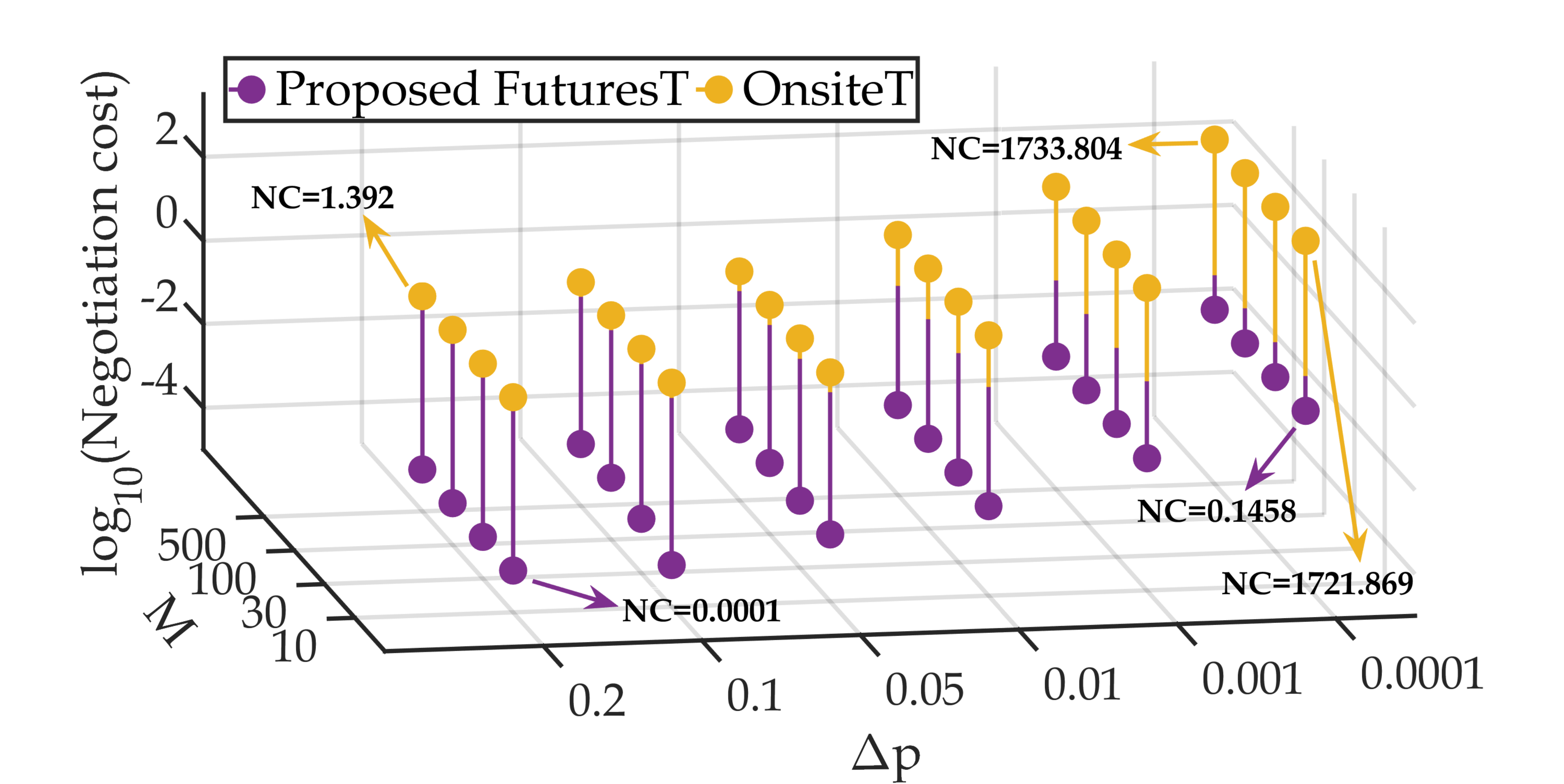}}\hfill
\subfigure[]{\includegraphics[width=.3\linewidth]{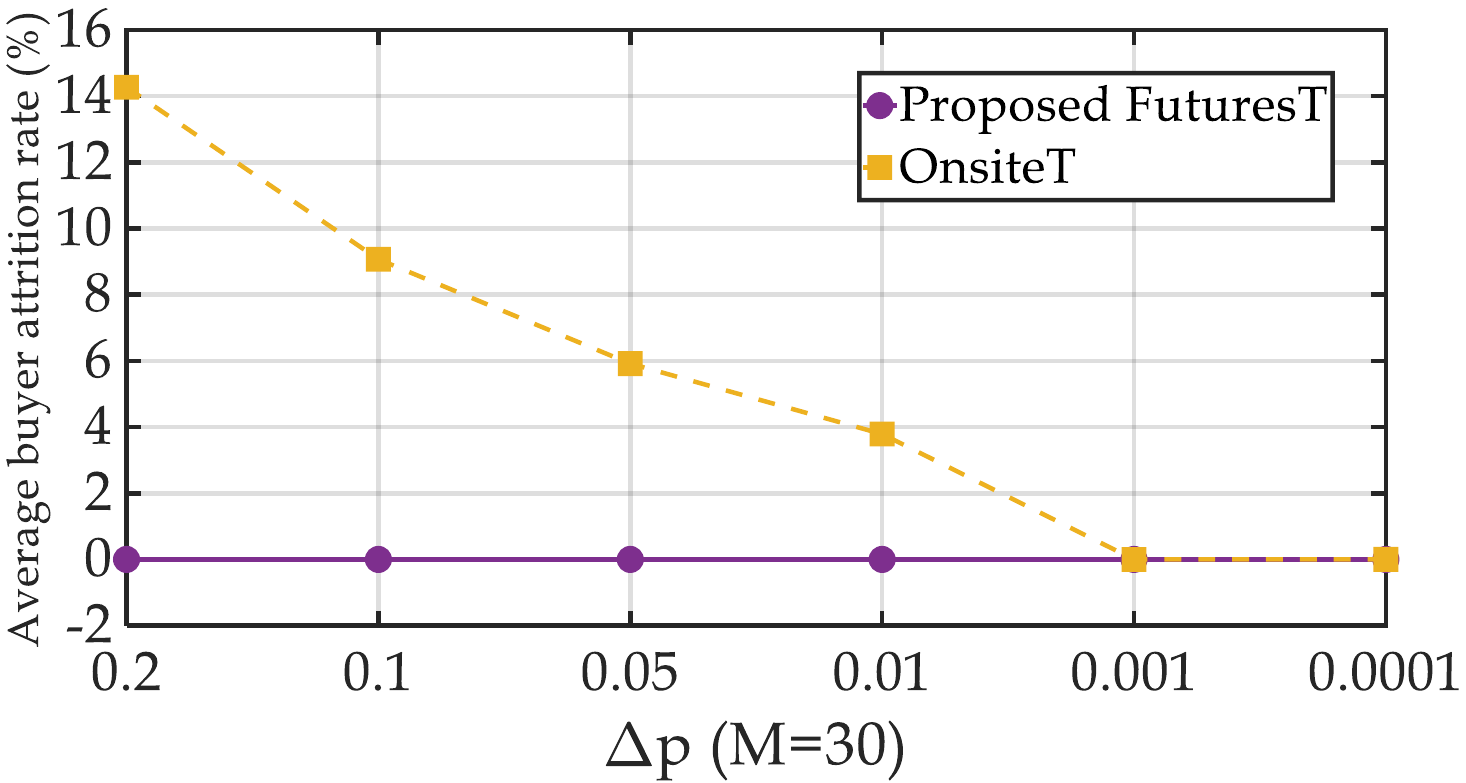}}\\
\subfigure[]{\includegraphics[width=.5\linewidth]{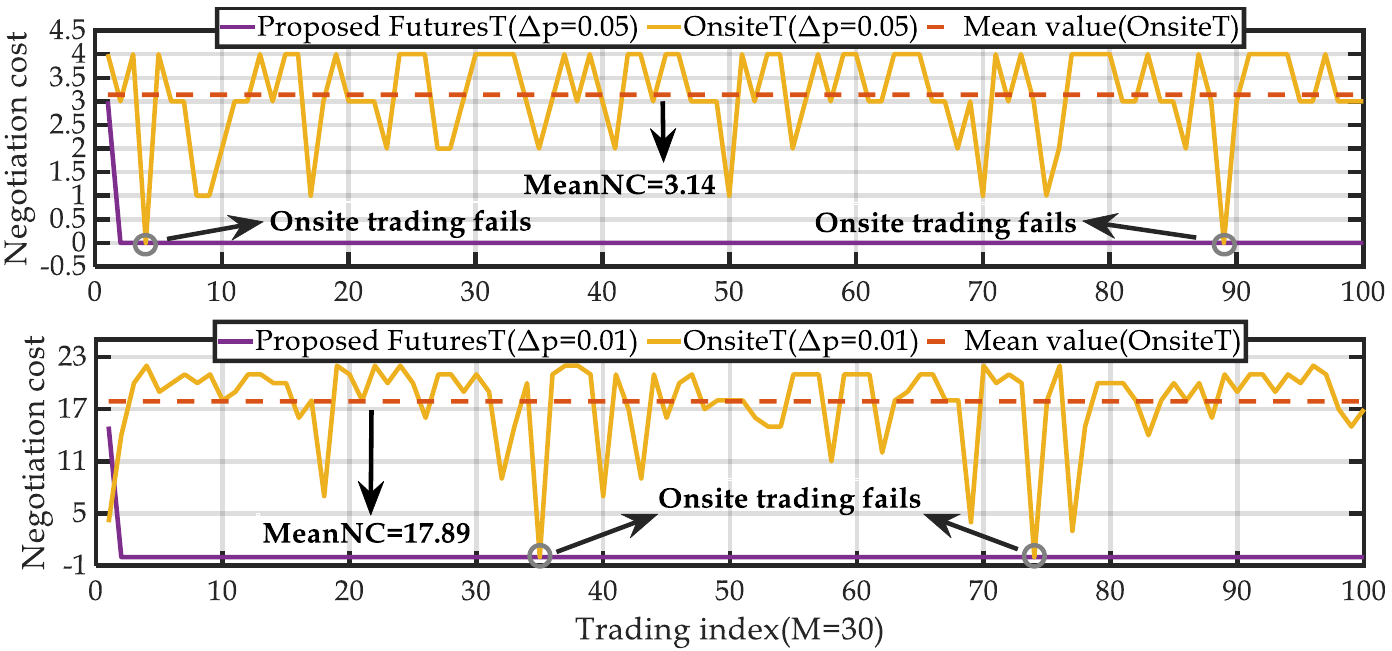}}\hfill
\subfigure[]{\includegraphics[width=.48\linewidth]{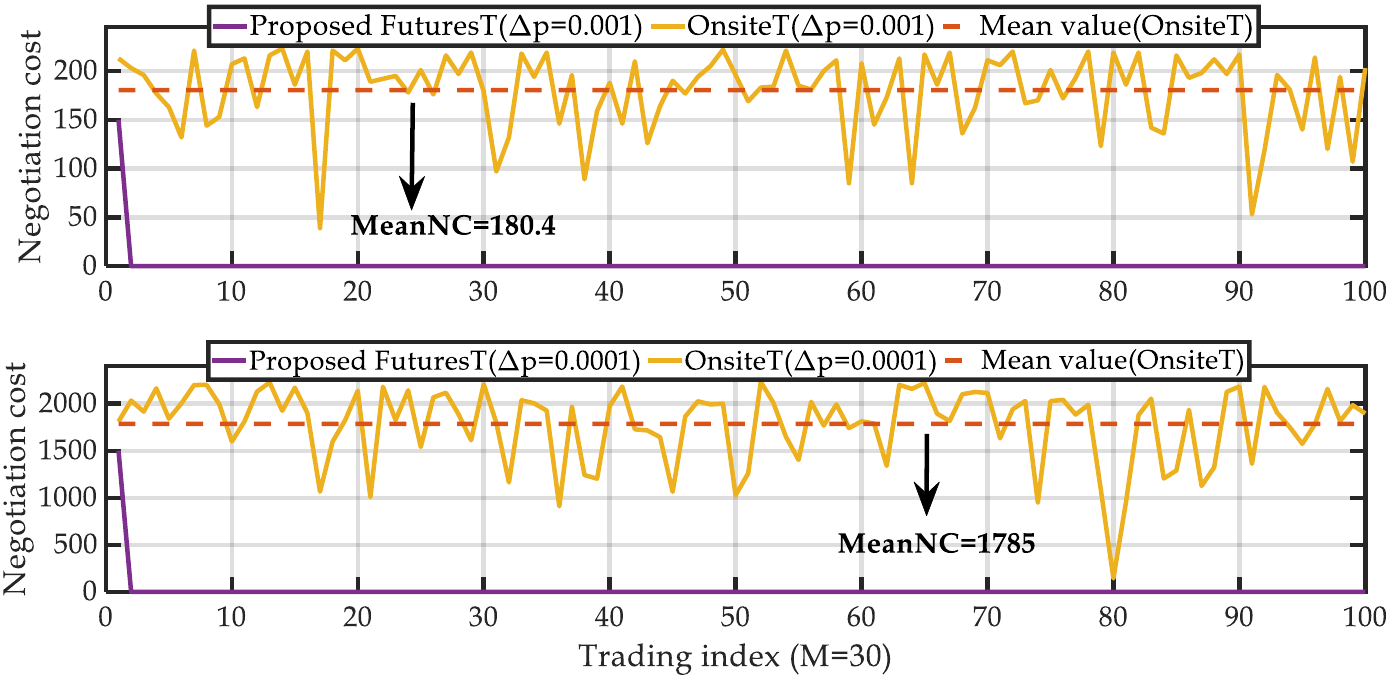}}
\caption{Performance comparisons upon having various $\Delta p$ and $M$.}
\end{figure*}

Apparently, a single trading can hardly reflect the advantage of the proposed FuturesT approach due to the unpredictability on the network conditions of a moving vehicle. Correspondingly, Fig.~5 demonstrates the performance evaluation and comparisons of the long-term participants' utilities, trading failures and the negotiation cost, where the cubic fitting curves are applied to better illustrate the long-term trend of each method. 

As can be seen from Fig.~5(a), the proposed FuturesT approach always outperforms the other methods on the sum utility of buyer with increasing number of trading. Fig.~5(b) demonstrates that the proposed approach achieves the similar performance on the sum utility of seller with OnsiteT. Moreover, the proposed FuturesT offers superior TFail, ABAR, NC and TFair rather than the OnsiteT method, as mentioned in Table~II. The seller's long-term utility of the FuturesT-noRE method keeps larger than the other methods but fails to protect the buyer's utility, which, however, brings difficulties to implement in real-life resource trading environments. Fig.~5(c) and Fig.~5(d) present the performance comparisons of TFail and NC of the OnsiteT method. Observed from Fig.~5(c), the value of TFail of OnsiteT increases upon rising the number of trading. Furthermore, the negotiation during each trading leads to higher NC, which is adverse to the long-term development of the resource trading market, as depicted in Fig.~5(d).

Figure~6 considers 100 trading to investigate the performance of participants' utilities affected by the number of local users $n_l$ and the SNR ${\gamma}_{b,s}$ of V2I communications. Fig.~6(a) and Fig.~6(b) present the utility of buyer and seller in each trading, respectively; Fig.~6(g) and Fig.~6(h) show the comparisons of price fluctuation and negotiation cost among different methods. Similar with Fig.~4, the proposed FuturesT gets better performance on the sum utility of buyer than the other methods (177.2440 for Proposed FuturesT, 115.2486 for OnsiteT, and $-$42.1843 for FuturesT-noRE); and achieves a commendable performance on the sum utility of seller (1457.67 for proposed FuturesT, 1708.08 for OnsiteT, and 2010.83 for FuturesT-noRE). 

Figure~6(c), Fig.~6(e), Fig.~6(d), and Fig.~6(f) illustrate the average utility of buyer and seller (per trading) upon having various $n_l$ and ${\gamma}_{b,s}$. For example, the average utility of buyer when $n_l=10$ in Fig.~6(c), is calculated as the sum utility of all the trading (when $n_l=10$) divides the total number of trading (when $n_l=10$). As shown in Fig.~6(c), the proposed approach always outperforms the FuturesT-noRE method, and achieves better performance than OnsiteT in most cases. In Fig.~6(d), the gradient of the curve of the proposed FuturesT approach ($\mathcal{A}^*=19$) reduces after $n_l=6$ owing to the positive waiting cost of local users, as depicted in Fig.~3(a). Fig.~6(e) shows that the larger values of SNR bring the buyer of the future-based trading with better average utilities. However, the curve of OnsiteT experiences a decline after ${\gamma}_{b, s}\in[18,20]$ dB due to that the buyer may have to bear a higher price. In Fig.~6(f), the average seller's utility of the onsite-based method rises with growing SNR ${\gamma}_{b, s}$ due to factors such like a higher price, where a larger value of ${\gamma}_{b, s}$ brings a lower ABAR. Moreover, the related performance of the two futures-based methods remain stable benefitting from the forward contract.

Figure~7 demonstrates the comparisons between the proposed FuturesT and the OnsiteT methods, upon having different values of $M$ and $\Delta p$. Specifically, Fig.~7(a), Fig.~7(b), and Fig.~7(c) investigates the performance evaluation on the average values of NL, NC and ABAR, via 10000 simulations (trading). Notably, we apply the 10-based logarithm representation in Fig.~7(a) and Fig.~7(b) since the gap between the two methods is too large. Overall, the proposed approach greatly outperforms the OnsiteT method on NL, NC and ABAR under any $M$ and $\Delta p$.
Specifically, the growing $M$ has a greater impact of NL on the proposed approach owing to the larger decision time of the buyer (line 7, Algorithm 1), as observed from Fig.~7(a). For example, the NL of the proposed FuturesT approach presents a nearly fourfold increase from $M=10$ to $M=500$, rather than around twofold increase of the OnsiteT method. Apparently, smaller $\Delta p$ will bring greater growth of NL and NC during negotiations. For instance, the NL of OnsiteT rises around 173~times from $\Delta p=0.2$ to $\Delta p=0.0001$ ($M=500$); and that of the proposed FuturesT has more than 2597-fold increase from $\Delta p=0.2$ to $\Delta p=0.0001$ ($M=10$). According to Fig.~7(b), the indicator NC is less affected by different $M$, but dramatically impacted upon changing the value of $\Delta p$. Consider $M=10$, the value of NC of OnsiteT rises from 1.392 per trading to 1733.804 per trading by reducing the value of $\Delta p$ from $0.2$ to $0.0001$, bringing tremendous overhead to the market in wireless network environment. On the contrary, according to Fig.~7(c), a smaller $\Delta p$ enables a lower ABAR, which, however, comes at the expense of larger NC and NL. As a result, the OnsiteT method can hardly achieve the tradeoff among NC, NL and ABAR. 

Moreover, Fig.~7(d) and Fig.~7(e) show the fluctuation of NC considering 100 trading and various $\Delta p$, which also well demonstrate the above-mentioned analytics.

\section{Conclusion}

Motivated by the challenges of heavy negotiation cost, unfairness, and trading failures in onsite resource trading environment, in this paper, we study a novel futures-based resource trading approach under EC-IoV. Specifically, the MEC sever with available computational resources, and the smart vehicle with heavy workload reach a consensus on the reasonable amount and price of resource for a forward contract, which will be fulfilled in the future. Aiming to maximize the expected utilities of both participants while evaluating the risks of possible losses, we propose an efficient bilateral negotiation mechanism. Simulation results demonstrate that the proposed futures-based resource trading approach can always achieve mutually beneficial situations for both the seller and buyer, while outperforming the baseline methods on critical factors such as the trading failures, the average buyer attrition rate, the negotiation latency and cost, as well as the trading fairness.

\clearpage

\appendix




\textbf{1. Derivation of ${\rm E}[C^{s}]$}

\noindent
According to (2), the expectation of $C^{\rm s}$ is given by (23).
\begin{align}
{\rm E}[C^{s}]& =\sum^M_{M-\mathcal{A}+1}{\frac{1}{M+1}({c_ln}_l-c_l(M-\mathcal{A}))}+\sum^{M-\mathcal{A}}_0{\frac{1\times 0}{M+1}}\notag \\
& =\sum^M_{M-\mathcal{A}+1}\frac{1}{M+1}({c_ln}_l-c_l(M-\mathcal{A}))\notag \\
& =\frac{c_l}{M+1}\sum^M_{M-\mathcal{A}+1}{n_l-\frac{\mathcal{A}c_l(M-\mathcal{A})}{M+1}}= \frac{{c_l\mathcal{A}}^2+c_l\mathcal{A}}{2 (M+1)}\tag{23}
\end{align}

\vspace{0.1in}

\noindent
\textbf{2. Derivation of (10) and (11)}

\noindent
Let discrete random variables $S_{1}$ and $S_{2}$ be,
\begin{align}
\label{eq24}
S_{1}& =n_lp_l,~~n_l\in \{0,p_l,2p_l, \cdots , p_l(M-\mathcal{A})\},\tag{24} \\
\label{eq25}
S_{2}& =(p_l-c_l)n_l+c_l(M-\mathcal{A}),~~n_l\in \{M-\mathcal{A}+1,\notag \\& \quad M-\mathcal{A}+2, \cdots, M\},\tag{25} 
\end{align}

Thus, we have
\begin{align}
\label{eq26}
{{\rm Pr} (S_{1}=k)} & =\frac{1}{M+1}, k\in \{0,1,\cdots, p_l(M-\mathcal{A})\}, \tag{26} \\
\label{eq27}
{{\rm Pr} (S_{2}=k')} & =\frac{1}{M+1}, k'\in \{p_l(M-\mathcal{A})+(p_l-c_l),\notag \\& \quad p_l(M-\mathcal{A})+2(p_l-c_l), \cdots, p_l M-c_l\mathcal{A}\}.\tag{27} 
\end{align}

Apparently, $S_{1}\neq S_{2}$, then (10) can be proved. According to the PDF of $ S$, the risk of seller can be calculated by considering the CDF of $S$, and (11) is thus proved. 
\vspace{0.1in}

%
%

\noindent
\textbf{3. Derivation of (18)}

\noindent
Let random variable $ Y=1+{\gamma}_{b, s}$, and $\dfrac{1}{Z}=\log_{2}(Y)$. Apparently, $Z$ represents a continuous random variable with the domain of definition $\left[\dfrac{1}{\log_{2}({\varepsilon}_{2}+1)}, \dfrac{1}{\log_{2}({\varepsilon}_{1}+1)}\right]$. Thus, the CDF of $Z$ can be calculated as (28):
\begin{align}
& {\rm F}_Z(z)={\rm Pr}(Z\le z)={\rm Pr}\left(\frac{1}{\log_{2}(Y)}\le z\right)={\rm Pr}\left(\log_{2}(Y)\ge \frac{1}{z}\right)\notag \\
&={\rm Pr}(Y\ge 2^{{1}/{z}})=1-{\rm Pr}(Y\le 2^{{1}/{z}})\notag \\
&
\begin{cases}
0, & z<\dfrac{1}{\log_{2}({\varepsilon}_2+1)} \\
1-\dfrac{2^{{1}/{z}}-{\varepsilon}_1-1}{{\varepsilon}_2-{\varepsilon}_1}, & \dfrac{1}{\log_{2}({\varepsilon}_2+1)}\le z\le \dfrac{1}{\log_{2}({\varepsilon}_1+1)} \\
1, & z>\dfrac{1}{\log_{2}({\varepsilon}_1+1)}.\tag{28} 
\end{cases}
\end{align}

Thus, we have the PDF of $Z$ in interval $\left[\dfrac{1}{\log_{2}({\varepsilon}_{2}+1)}\right.$, $\left.\dfrac{1}{\log_{2}({\varepsilon}_{1}+1)}\right]$ as (29), where $r''=\dfrac{\ln 2}{{\varepsilon}_2-{\varepsilon}_1}$ denotes a constant for notational simplicity.
\begin{align}
\label{eq31}
& {\rm Pr}(Z=z)=\frac{\partial {\rm F}_Z(z)}{\partial z}=r''\times \frac{2^{{1}/{z}}}{z^2},\notag \\
&\qquad z\in \left[\frac{1}{\log_{2}({\varepsilon}_{2}+1)}, \frac{1}{\log_{2}({\varepsilon}_{1}+1)}\right]\tag{29} 
\end{align}

According to (29), we calculate ${\rm E}\left[\dfrac{1}{\log_2(1+{\gamma}_{b, s})}\right]$ as below:
\begin{align}
& {\rm E}\left[\frac{1}{\log_2(1+{\gamma}_{b, s})}\right]={\rm E}[Z]\notag \\
& =\int^{\frac{1}{{l{\text{og}}}_{2}({\varepsilon}_{1}+1)}}_{\frac{1}{\log_{2}({\varepsilon}_{2}+1)}}{z{\rm Pr}(Z=z){d}z}=r''\int^{\frac{1}{\log_{2}({\varepsilon}_{1}+1)}}_{\frac{1}{\log_{2}({\varepsilon}_{2}+1)}}{\left(\frac{2^{{1}/{z}}}{z}\right){d}z}\notag \\
\label{eq32}
& =r''{\rm Ei}({\ln 2}\times \log_{2}({\varepsilon}_{2}+1))-r''{\rm Ei}({\ln 2}\times \log_{2}({\varepsilon}_{1}+1)),\tag{30} 
\end{align}
where $\left(\dfrac{2^{{1}/{z}}}{z}\right)$ stands for a nonelementary function which poses challenges on definite integral. Thus, we apply exponential integral $\rm Ei ()$ to represent (30). Correspondingly, (18) is proved. \\

\vfill

\end{document}